\documentclass[journal]{IEEEtran}

\ifCLASSINFOpdf
\else
   \usepackage[dvips]{graphicx}
\fi
\usepackage{url}
\usepackage{amsmath}
\usepackage{bm} 
\usepackage{nicefrac}
\usepackage{amssymb}
\usepackage{stmaryrd}
\usepackage{upgreek}
\usepackage[T1]{fontenc}
\usepackage{comment}
\usepackage{graphicx}

\begin{document}

\title{Event-based Processing of Single Photon Avalanche Diode Sensors}

\author{Saeed Afshar,~\IEEEmembership{Member,~IEEE}, Tara Julia Hamilton,~\IEEEmembership{Member,~IEEE}, Langdon Davis, Andr\'e van Schaik,~\IEEEmembership{Fellow,~IEEE}, and Dennis Delic,~\IEEEmembership{Member,~IEEE}
\thanks{``This work was supported by” add “NATO Science for Peace and Security (SPS) Programme 98480.'' }

\thanks{S. Afshar, T. J. Hamilton, A. van~Schaik and D. Delic are with the International Centre for Neuromorphic Systems, Western Sydney University, Sydney Australia, 2747.}
\thanks{T. J. Hamilton is with the School of Engineering, Macquarie University, North Ryde, NSW 2109, Australia.}
\thanks{L. Davis is with BAE systems, Australia.}
\thanks{D. Delic is with Defence Science and Technology, Australia.}
\thanks{Corresponding author Email: s.afshar@westernsydney.edu.au.}}

\markboth{}
{Shell \MakeLowercase{\textit{et al.}}: Bare Demo of IEEEtran.cls for IEEE Journals}
\maketitle

\begin{abstract}

Single Photon Avalanche Diode sensor arrays operating in direct time of flight mode can perform 3D imaging using pulsed lasers. Operating at high frame rates, SPAD imagers typically generate large volumes of noisy and largely redundant spatio-temporal data. This results in communication bottlenecks and unnecessary data processing. In this work, we propose a set of neuromorphic processing solutions to this problem. By processing the SPAD generated spatio-temporal patterns locally and in an event-based manner, the proposed methods reduce the size of output data transmitted from the sensor by orders of magnitude while increasing the utility of the output data in the context of challenging recognition tasks. To demonstrate these results, the first large scale complex SPAD imaging dataset is presented involving high-speed view-invariant recognition of airplanes with background clutter. The frame-based SPAD imaging dataset is converted via several alternative methods into event-based data streams and processed using a range of feature extractor networks and pooling methods. The results of the event-based processing methods are compared to processing the original frame-based dataset via frame-based but otherwise identical architectures. The results show the event-based methods are superior to the frame-based approach both in terms of classification accuracy and output data-rate. 
\end{abstract}


\IEEEpeerreviewmaketitle

\section{Introduction} \label{sec:intro}
\subsection{Single Phone Avalanche Diodes Sensor}\label{sec:intro/SPAD}

\IEEEPARstart{A} Single Photon Avalance Diode (SPAD) is a type of photo-detector that comprises of a reversed biased photo-diode operated just below the breakdown voltage and as such is able to detect individual incoming photons from the environment \cite{zappa2007principles}. This ability to detect single photons enables SPAD cells to calculate precise photon timing information. Integrating an array of SPAD detectors onto a single CMOS chip and using high precision laser illuminators allows the development of SPAD cameras, which can capture high-speed 3D images under extremely low-light conditions. SPAD array cameras have a broad range of applications from military, meteorology, space, augmented reality, remote sensing, autonomous robotics \cite{zappa2005spada}\cite{bellisai2013single}.
SPAD imagers can operate in two distinct modes. As shown in Figure~\ref{fig:SpadAppraoch} (a), in the photon timing, or Direct Time Of Flight (DTOF) mode, each pixel uses a high speed counter to measure the time interval it takes for photons from an illuminating laser pulse to reach the object being imaged and return to the sensor. By measuring the time it takes for light to travel to the target and back the imager captures a 3D image of the environment. The second mode of operation for the SPAD imager is photon counting or Indirect Time Of Flight (ITOF). Here the imager counts number of returning photons from the illuminating laser or background illumination. In this mode, the sensor operates in a similar manner to a conventional imager with very high sensitivity that is capable of detecting single photons.

\subsection{Event-based Processing and SPAD}
\label{sec:SPAD/intro/EB}

The conventional approach to date has been to encode the time of flight of the arriving photons using high precision counters for each SPAD cell and to transfer this timing data off-chip for processing~\cite{zappa2007principles}. This approach typically involves as a first step some form of averaging over a large number of frames which effectively removes the possibility of on-chip processing. This transfer process also creates an information bottleneck which is currently one of the major limiting factors in the speed of operation of high frame rate SPAD cameras. In addition, the use of conventional CPUs or GPUs for processing this temporal data makes processing SPAD data computationally intensive using conventional signal processing techniques and results in significant power and hardware requirements.

Yet the attributes that make SPAD data challenging for conventional processors, when combined with the significant level of temporal redundancy present in real-world visual data, makes the SPAD cell activation patterns ideal for event-based and spiking neuromorphic processors that are designed to operate directly on noisy temporal data in a parallel fashion. 

While the conversion of high data-rate DTOF SPAD data into local event-based features is entirely novel, previous works have demonstrated the utility of taking bio-inspired approach to SPAD processing. In \cite{berkovich2015scalable}, Berkovich et al. present a scalable 20$\times$20 SPAD imaging array using asynchronous Address Event Representation (AER) readout. In \cite{shamim2018cmos} the same approach is proposed for use in Positron Emission Tomography applications. The AER protocol is an efficient communication protocol for sparse event-based data which reports events as they occur removing the need for global frames \cite{boahen1998communicating}. In these works, the SPAD cells operate in an ITOF photon counting mode where an analog photon-counting circuit counts incoming photon until the counter reaches a preset threshold causing the  pixel to generate an event indicating a preset level of illumination. This mode of operation is similar to previously proposed non-SPAD event-based sensors \cite{lichtsteiner2008128} albeit with the advantage of the SPAD's high quantum efficiency. In contrast, the design proposed in this work seeks to combine the inherently temporal nature of DTOF SPAD spatio-temporal data with neuromorphic event-based feature extraction and processing.

In the proposed approach instead of encoding, storing and transferring the high-resolution, (typically 16 bit) photon time of flight data off-chip for processing, the measurement of the time of flight of the laser pulse is abandoned entirely in favor of a neuromorphic processor that operates directly in the time domain and on the inter-spike intervals within local regions of the SPAD array. The proposed approach illustrated in Figure \ref{fig:SpadAppraoch} motivates the development and hardware implementation of event-based feature extraction algorithms and circuits that generate sparse event-based local representations from the non-sparse event-based SPAD activation data and in this way drastically reduce the I/O requirements of the overall system. 

\section{Methodology}
\label{sec:SPAD/meth}
\subsection{The SPAD Dataset}
\label{sec:SPAD/meth/dataset}

\begin{figure}
 \centering
 \includegraphics[width=.4\textwidth]{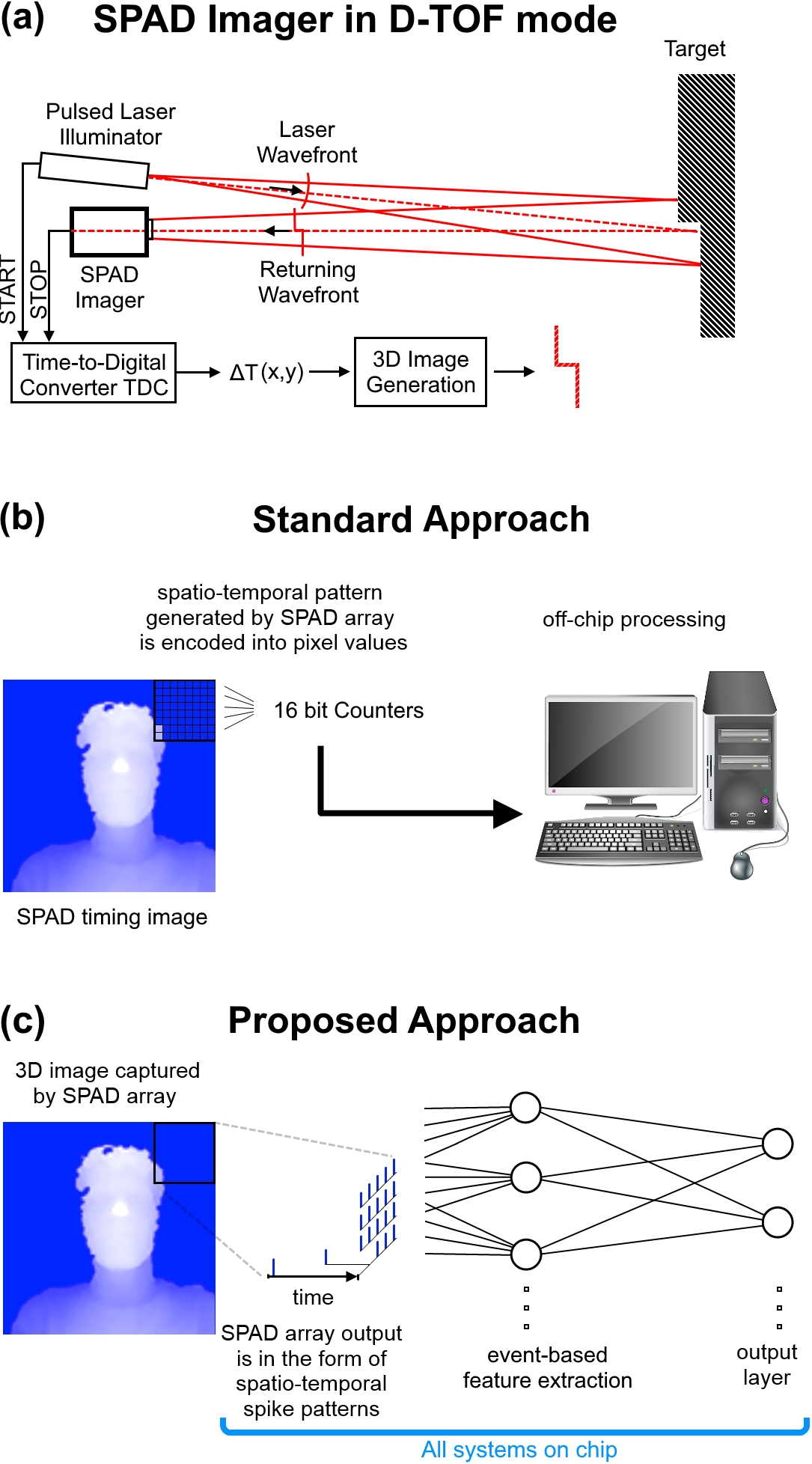}
 \caption{\textbf{Conventional and Neuromorphic SPAD DTOF data processing.} (a) SPAD imager in DTOF mode using a pulsed laser illuminator. Using SPAD sensors in Direct-Time of Flight (DTOF) mode enables the capture of three-dimensional images with a single camera. (b) Standard approach to processing SPAD imaging data using on-chip counters and off-chip processing. (c) Proposed event-based approach to SPAD data processing.} 
 \label{fig:SpadAppraoch}
\end{figure}

In this work, we tackle the challenge of performing classification of a large complex SPAD imaging dataset generated using frame-based and event-based approaches. The task involves recognition of fast moving model airplanes. The view-invariant classification of the fifteen classes of target airplanes and one distractor represents a challenging problem given the similarity of the classes, low spatial resolution, presence of partial occlusions and the high noise level in the dataset.

\begin{figure}
 \centering
 \includegraphics[width=.5\textwidth]{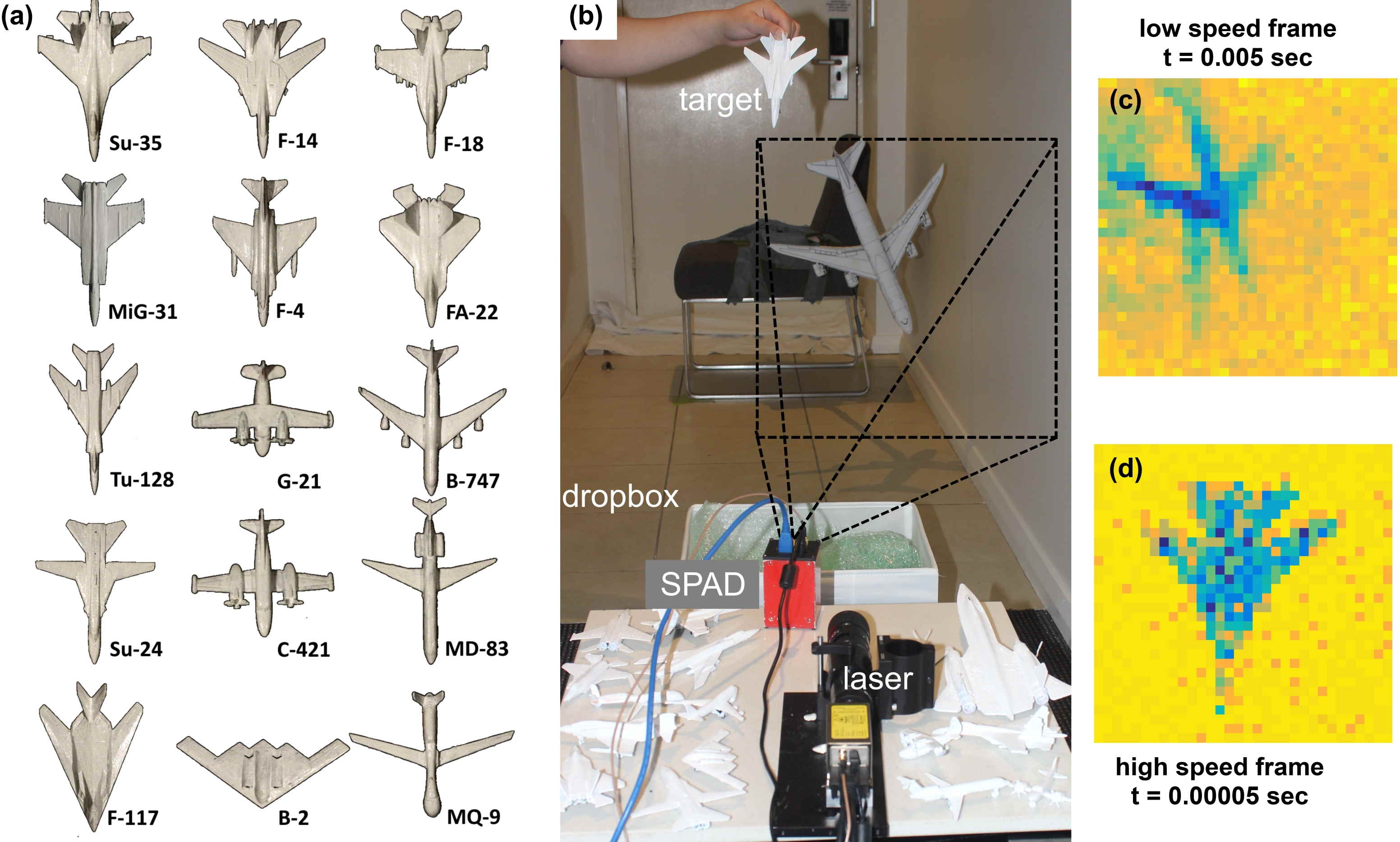}
 \caption{\textbf{SPAD sensor airplane drop classification experiment.} (a) Fifteen model airplane types make up the 15 classes in the detection and classification task. (b) Experiment set-up. Metallic model airplanes painted a uniform white were dropped in from of the SPAD sensor at close range (approximately 30-40cm) resulting in high relative velocity. SPAD field of view is marked by the black dotted line. In the background (approximately 3 meters) a large model B-747 airplane serves as a distractor. (c) SPAD image generated from averaging 500 raw frames representing 5~ms of recording time. The background B-747 model is clearly visible. (d) SPAD image showing of the rapidly moving F-14 model generated through averaging 5 raw frames representing 50~$\mu$s of recording time.}
 \label{fig:datasetSetup}
\end{figure}

The 32$\times$32 pixel SPAD camera used in this work was fabricated on a standard CMOS chip with each pixel integrating one SPAD and one time-to-digital converter as illustrated in Figure \ref{fig:SpadAppraoch}(a) and described in \cite{villa2014cmos}. The SPAD camera was fitted with a Navitar NMW-12WA lens and a Thorlabs 660 nm filter. The SPAD camera's field of view was set to 26.22 degrees. The laser used to obtain DTOF data was a 100 mW 660 nm Coherent CUBE diode laser using a 12$\times$ zoom lens such that the region of laser illumination and the SPAD camera field of view were overlayed as shown in Figure \ref{fig:datasetSetup}(b). 

The targets in the dataset are imaged using the SPAD sensor in a photon timing mode where each SPAD pixel operates as a LIDAR sensor. The illuminating laser is pulsed at 100 kHz providing photon time of flight information at an extremely high frame rate. By dropping the model airplane at high speed close to the sensor, the high temporal resolution of the sensor can be leveraged and investigated. As shown in Figure \ref{fig:datasetSetup}(b), the experiment involves the use of a larger more distant background stationary B-747 model as a distractor. This distractor becomes increasingly more prominent as the number of frames collected for an image is increased. The inclusion of the larger stationary distractor with the high-speed target ensures that the dataset can only be processed at high speed precluding the option of increasing SNR via frame averaging. This set-up ensures a high noise imaging signal that better represents real-world imaging environments. Unlike controlled image collection environments typically used in machine vision research, real-world imaging environments are unpredictable, dynamic and noisy, precluding many commonly used image enhancement methods such as frame averaging. This experiment design aims to encourage the development of algorithms that are robust to noise and can more readily be applied to challenging real-world imaging environments.

\begin{figure*}
 \centering
 \includegraphics[width=.75\textwidth]{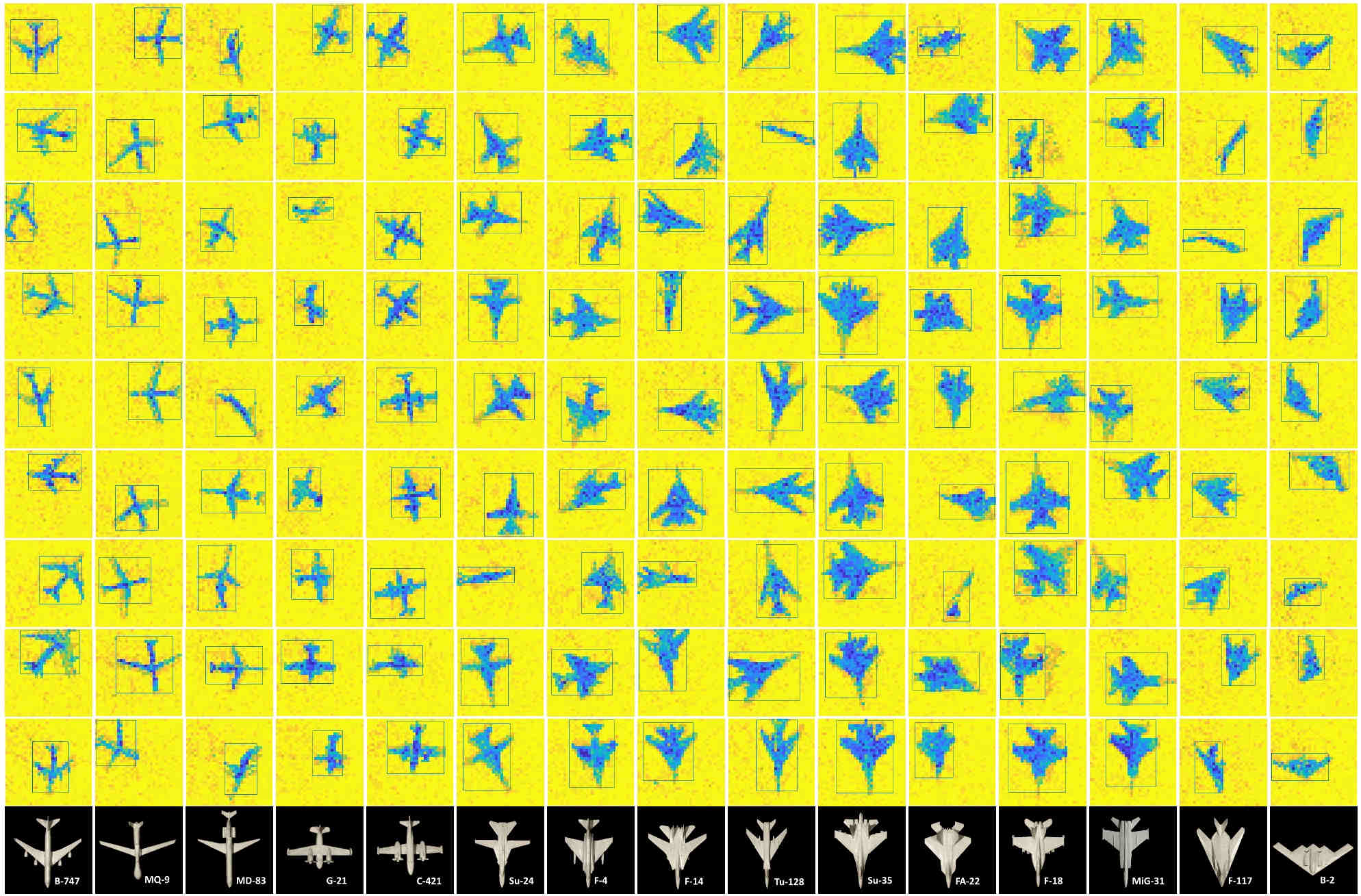}
 \caption{\textbf{Sample recordings from the SPAD dataset.} Each of the fifteen columns shows random samples of each airplane class in the dataset. The images show the wide range of observed orientations, sizes and partial occlusions as the model airplanes pass through the field of view. The bottom row shows a photo and label of each model. Note that the images show only the midpoint of the sample recordings.}
 \label{fig:datasExamples}
\end{figure*}

The originally captured dataset involved 3000 individual uncontrolled free hand drops of the 15 airplane classes with 200 drops per class. The dataset and associated supporting files are available for download at \cite{SPAD_Plane_Dataset2019}. This 3000 recording dataset was augmented via mirror reflection as well as 90, 180 and 270 degree rotation resulting in an augmented dataset of 24000 recordings. Sample recordings from the dataset are shown in \ref{fig:datasExamples} illustrating the significant visual complexity due to variance in target orientation, occlusions and the similarity of the tested classes. This complexity is even greater when the entire video of each recording is considered due to the change in relative orientation of each target during each recording and due to the occlusions present at the beginning and end of the recordings as the targets enter and exit the field view.
 
\subsection{First-AND Event Generation Method: Discarding Time and Transmitting Change}
\label{sec:SPAD/meth/frame2event/First-AND}

In previously implemented SPAD DTOF systems and in our proposed system, when a SPAD pixel is activated, it enables a latch which stays high until it is reset. The reset is typically performed after all data from the current laser pulse has been transmitted off the sensor. This data and the associated delay can be significant especially as the number of pixels on the imager becomes very large. In the first proposed system, which we call First-AND, instead of recording and transmitting the time interval from the initial laser pulse, only the inter-pixel photon arrival order (not the time) is detected. In this way, the requirement for precise measurement and transmission of the photon time of flight is removed along with the resource consuming high-precision, on-chip, per-pixel counters and memory circuits.

This simplification is achieved through the use of multiple AND gates which take as input a local group of pixels. The number of input pixels per AND gate must be equal so as to provide an equal probability of gate activation. The pattern of connectivity and its correlation to the observed spatio-temporal order of SPAD activation also determines AND gate activation. For example, an edge bar is more likely to be activated in a natural environment than a checkerboard pattern since the latter is not typically observed in the visual environment. In this way, each AND gate encodes a local feature and its activation indicates that all its input signals have been activated. The choice to use digital AND gates, as opposed to an analog summing and comparator circuits was made to simplify the Integrated Circuit design of the system and to ensure a deterministic output for each gate.

The AND gate pattern used in this work have overlapping receptive fields and are tiled across the visual spatial field to form a convolutional layer. Thus, the same pattern of AND gate connectivity is repeated across the visual field. Each AND gate can be interpreted as a neuron in a local single-layer network of $N_0$ (in this case $N_0$=4) neurons connected to a local  $r \times r$ (in this case $r = 4$) receptive field. For each receptive field as soon as all the input pixels of a single AND gate latch high, i.e. as soon as the all SPADs feeding an AND gate detect a photon, the AND gate goes high.

In the proposed design, the latching of the first AND gate at each receptive field at each laser pulse, prevents subsequent latching of any later neuron at that receptive field via a recurrent enable connection that gates all AND gates. This temporal inhibitory feedback structure was introduced in the SKAN network \cite{Afshar2014c} and demonstrated in FPGA hardware. 

In this approach, each receptive field only requires memory storage for a two bit address of the feature which was detected most recently. Furthermore, the design can be modified to only transmitting changes in feature detection at each receptive field, rather than reporting the winning feature each laser pulse. Given an ideal, noise free sensor, this modified method would greatly reduce the output data rate of the event-based sensor. Unfortunately, in practice the sensor noise can generate a significant amount of random feature changes resulting in a large number of noise events.

Based on the data recorded during our experiment, a range of sensor non-idealities and noise sources were found. These non-idealities could be broadly divided into false-positive and false-negative latching events, imprecise timing in the latching of the SPAD pixels (jitter) and persistent non-ideal timing patterns across the array pixels. These noise sources are detailed in the Supplementary Material.

Through consideration of the likely sources of noise, the event generation method of the First-AND system based on local changes in features is modified to introduce robustness to noise whereby a feature detection success counter is added such that every time a detected feature in a receptive field is the same as the one already in memory for that receptive field, the detection counter increments by one. Every time the newly detected feature is different from the one already in memory, the counter is decremented by one. If the feature detection counter of a receptive field reaches a pre-set threshold value ($\Phi = 6$), the receptive field creates a feature event and the counter is set to zero. Conversely, if the counter reaches zero after a decrement, the newly detected feature which caused the decrement replaces the old feature that was in memory. In this way, a constantly noiselessly detected feature will periodically send out a confirmatory feature event, whereas receptive fields where no feature consistently wins will not output any features. By decreasing the global threshold, we can decrease the number of times a feature must be detected before it triggers a feature event. This reduction in threshold increases the data-rate and allows features whose verity is less certain to be transmitted. Conversely, a higher threshold increases the certainty about the transmitted features and reduces the data-rate. In this way a global feedback control to the system can be implemented. While this aspect of the design was not explored it forms an avenue of investigation in future work.

\subsection{Implementation of the First-AND Network in ASIC}
\label{sec:SPAD/meth/frame2event/AsicSpad}
The First-AND system described in Section \ref{sec:SPAD/meth/frame2event/First-AND} was implemented as part of a Complementary Metal-Oxide-Semiconductor (CMOS) based SPAD cells with supporting mixed-signal and digital Integrated Circuit (IC) chip design using a Silterra High Voltage CL130H32 Process Technology.
The chip operates in time of flight mode, in this configuration the start of a new acquisition cycle is synchronized to a laser pulse being fired at the target (i.e. Flash LADAR). The implemented $128 \times 128$ SPAD array contained $125\times 125$ $4\times 4$ receptive fields each with 4 silicon digital AND based feature detectors consisting of North, South, East and West. The system features priority encoding, a 3-bit feature counter and an adjustable threshold for the detection of winning features. The network readout is implemented via the AER protocol allowing an asynchronous DTOF SPAD sensor array with real-time event-based feature extraction for 3D imaging applications. The inclusion of the 3-bit feature success counter and a globally adjustable feature count threshold provides robustness to noise while minimizing the readout of noise from the sensor.

\begin{figure*}
 \centering
 \includegraphics[width=1\textwidth]{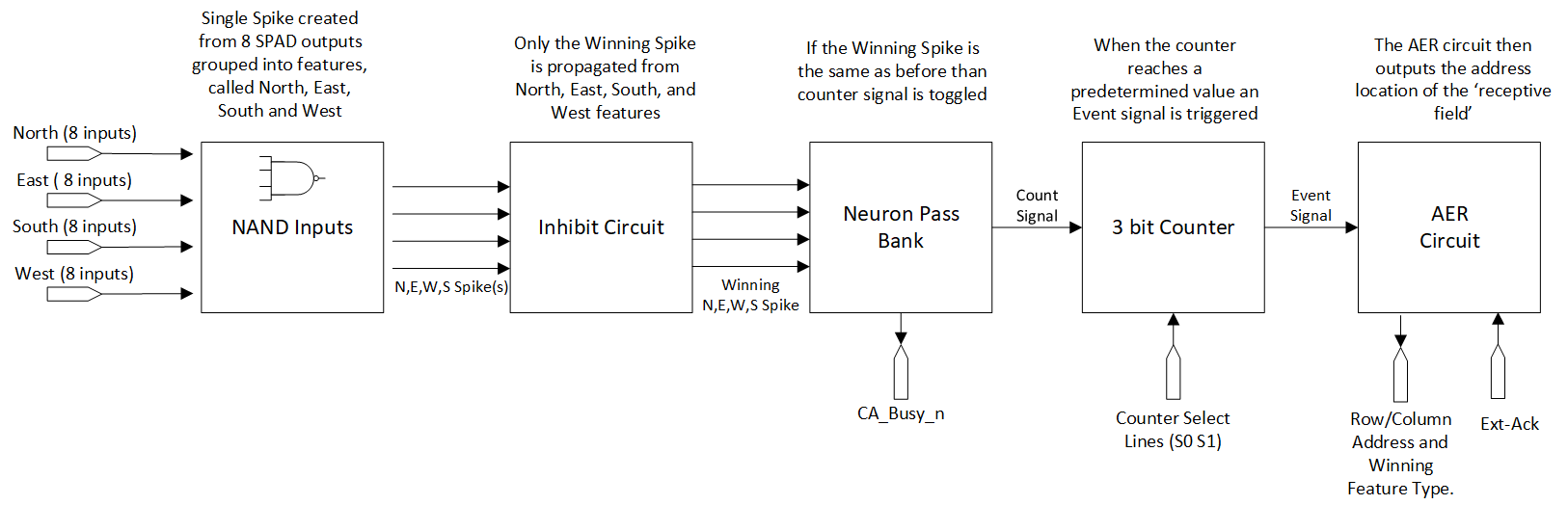}
 \caption{\textbf{Functional block diagram of a single receptive field of the ASIC implemented First-AND Network.}}
 \label{fig:AsicBlockDiagram}
\end{figure*}

Figure \ref{fig:AsicBlockDiagram} shows a functional representation of each receptive field circuit or cell.  The 16 SPADs connections are not shown. When the SPADs avalanche, or fire, they are synchronously latched to an on-chip CLOCK (configured via PLL or fed via external CLOCK signal). The value of the threshold $\Phi$, is set by the input S0/S1 lines shown in Figure \ref{fig:AsicBlockDiagram}. The event is indicated by the CA\_Busy\_n output line on the chip. When an event is generated, the encoded row and column address of the location of this receptive field is sent off-chip as well as the class of winning feature North, South, East or West. When the user has read the address/data information, an acknowledgement is sent to the ext-ack input line which resets the counter and releases the CA\_Busy\_n line. The event generator in a particular receptive field is reset once it receives row and column acknowledge signals as well as a global acknowledge signal which is generated off-chip. Although the acquisition cycle is synchronized to CLOCK, events are asynchronously generated off the chip. The on-chip arbiter processes the order of events as they occur, and events can be generated asynchronously, as such it is possible that not all events will be captured and read by the monitoring FPGA between laser pulses resulting in dropped events. 

\begin{figure}
 \centering
 \includegraphics[width=.45\textwidth]{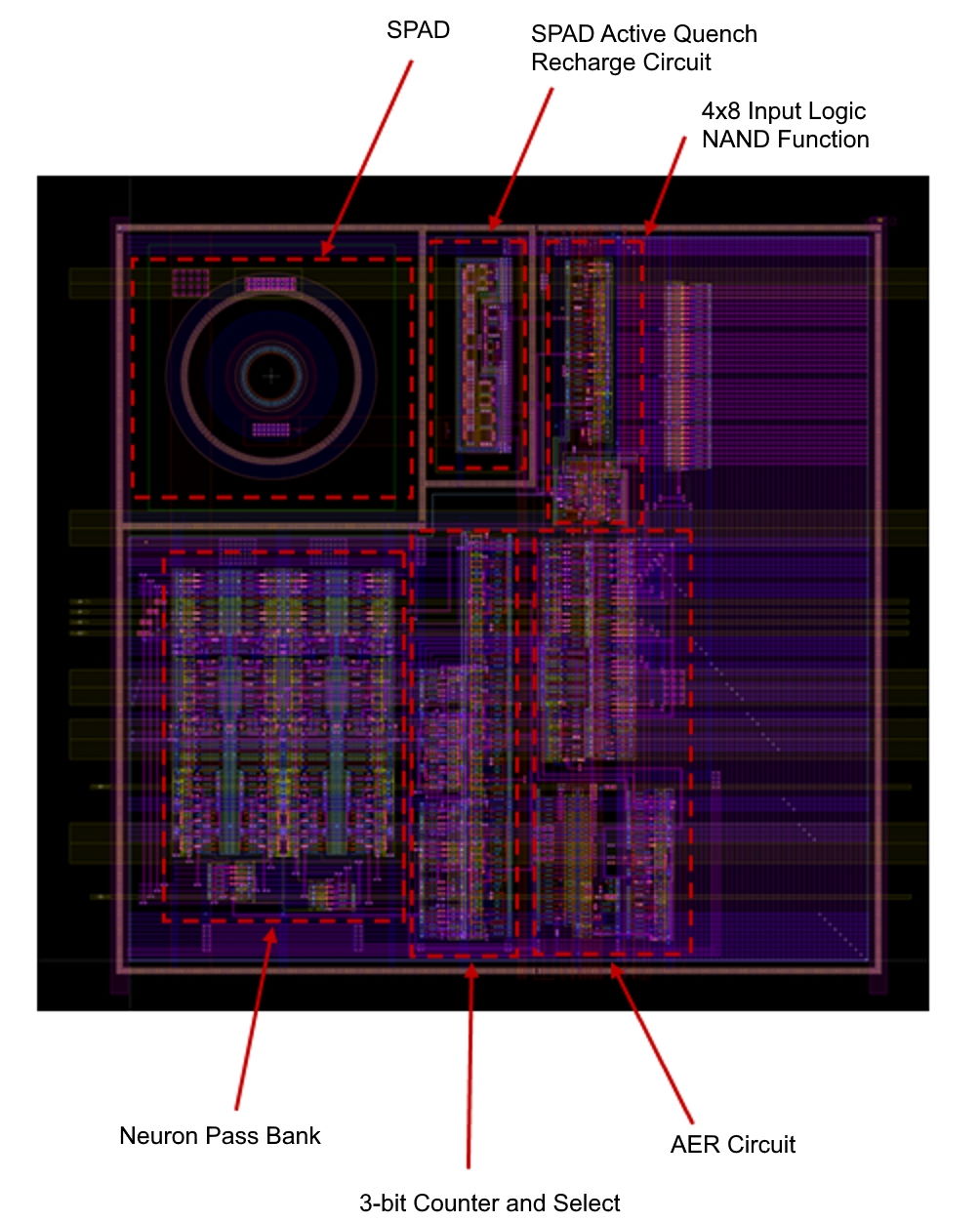}
 \caption{\textbf{Single Receptive Field Cell showing a SPAD sensor and Circuitry Blocks, size 75 $\bm\mu$m $\times$ 75 $\bm\mu$m. Here the neuron pass bank determines whether the current detected feature is the same as the previous detection.}}
 \label{fig:AsicSpadPixel}
\end{figure}

\begin{figure}
 \centering
 \includegraphics[width=.5\textwidth]{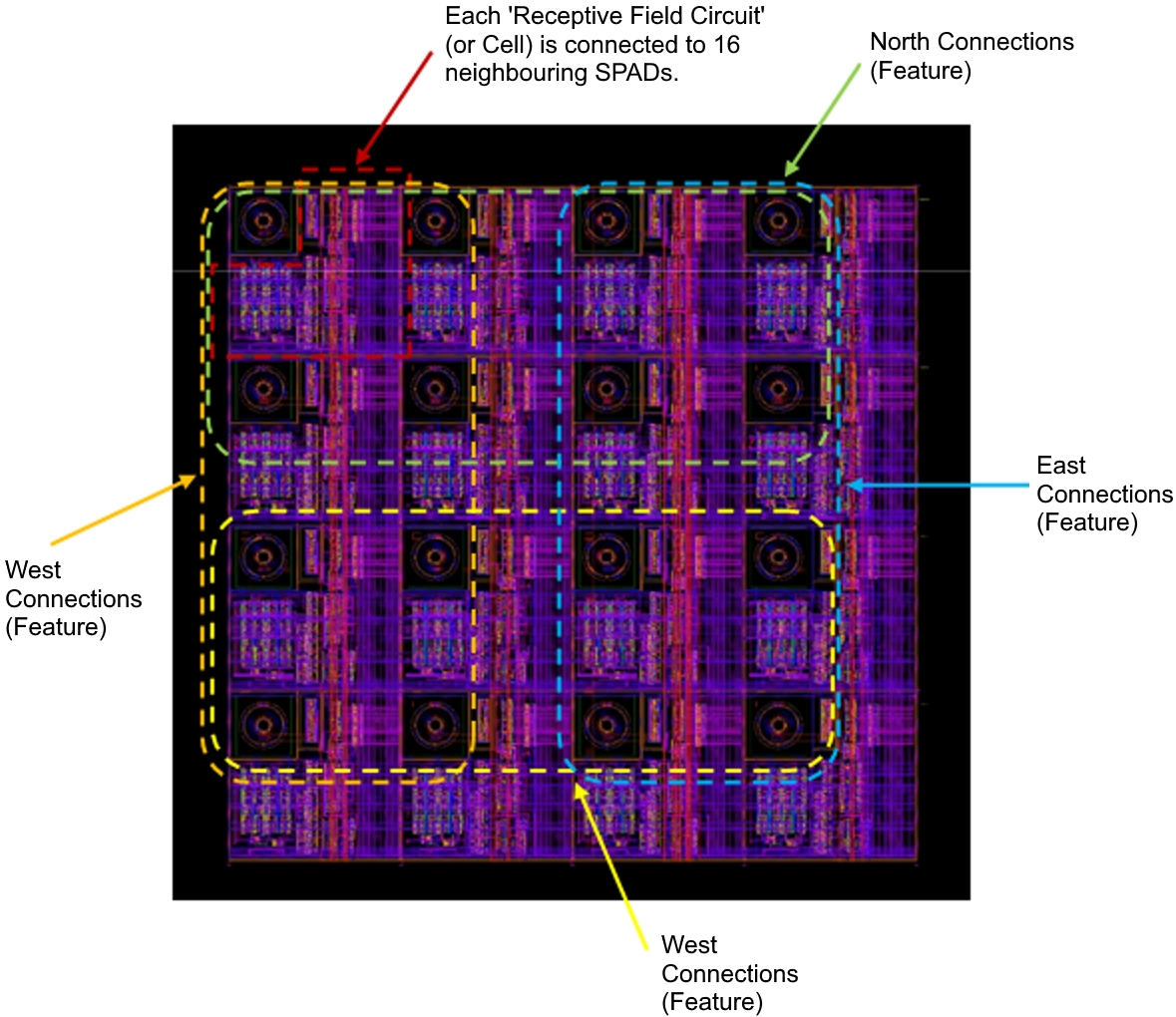}
 \caption{\textbf{Each receptive field is comprised of (connected to) 16 (4$\times$4) neighboring SPAD detectors 30$\bm\mu$m in diameter (5$\bm\mu$m active area).}}
 \label{fig:AsicSpadRf}
\end{figure}

Figures \ref{fig:AsicSpadPixel} and \ref{fig:AsicSpadRf} show the layout of a single SPAD pixel and a 4$\times$ 4 pixel receptive field respectively.

\subsection{Training Binary Feature Extractor Networks}
\label{sec:SPAD/meth/frame2event/feast}
In order to extract higher level spatio-temporal patterns generated by the SPAD imager, an event-based feature extraction network was trained on the event-based dataset. The Feature Extraction using Adaptive Selection Thresholds (FEAST) method used was detailed in \cite{afshar2019Feast}. This simple event-based unsupervised learning algorithm uses competition between adaptive neurons to generate balanced network activation in response to incoming data.

\begin{figure}
 \centering
 \includegraphics[width=.5\textwidth]{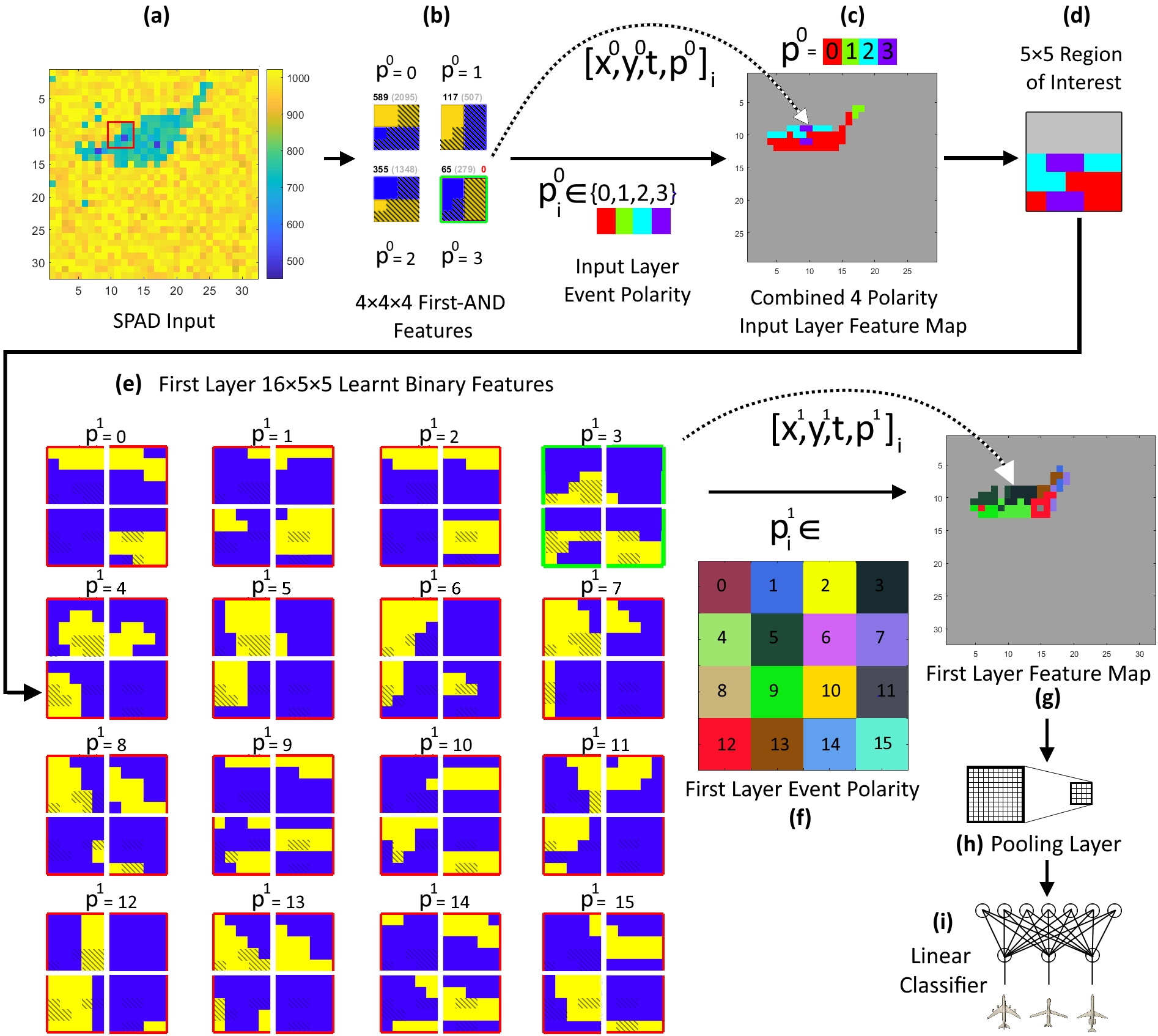}
 \caption{\textbf{Block diagram of the end-to-end First-AND event-based processing system.} (a) Shows the raw image generated by the SPAD sensor in time of flight mode as a B-2 model enters the field of view. The red box indicates the receptive field of the current generated event. (b) Shows the four First-AND features and their binary bar shaped weights. Superimposed are the state of the latched SPAD pixels at the moment the First-AND feature generates an event (diagonal black lines). The third feature is the first AND gate to latch disabling the others and passing its event to the next layer. (c) Shows $\bm S^0_i$, the binary-valued four-polarity time surface with activation over $\tau_0$ = 2~ms. This surface serves as a feature map for the next layer of processing. (d) Shows the 5$\times$5 Region of Interest ROI extracted from $\bm S^0_i$. (e) Shows the 16 four-polarity binary event-based features which operates on $\bm S^0_i$. (f) Shows the color coding of the 16 features. (g) Shows the 16 polarity binary-valued time surface $\bm S^1_i$. Panels (h) and (i) show the pooling and classification layers respectively.}
 \label{fig:neuroSpadBlockDiagram}
\end{figure}

\begin{figure}
 \centering
 \includegraphics[width=.5\textwidth]{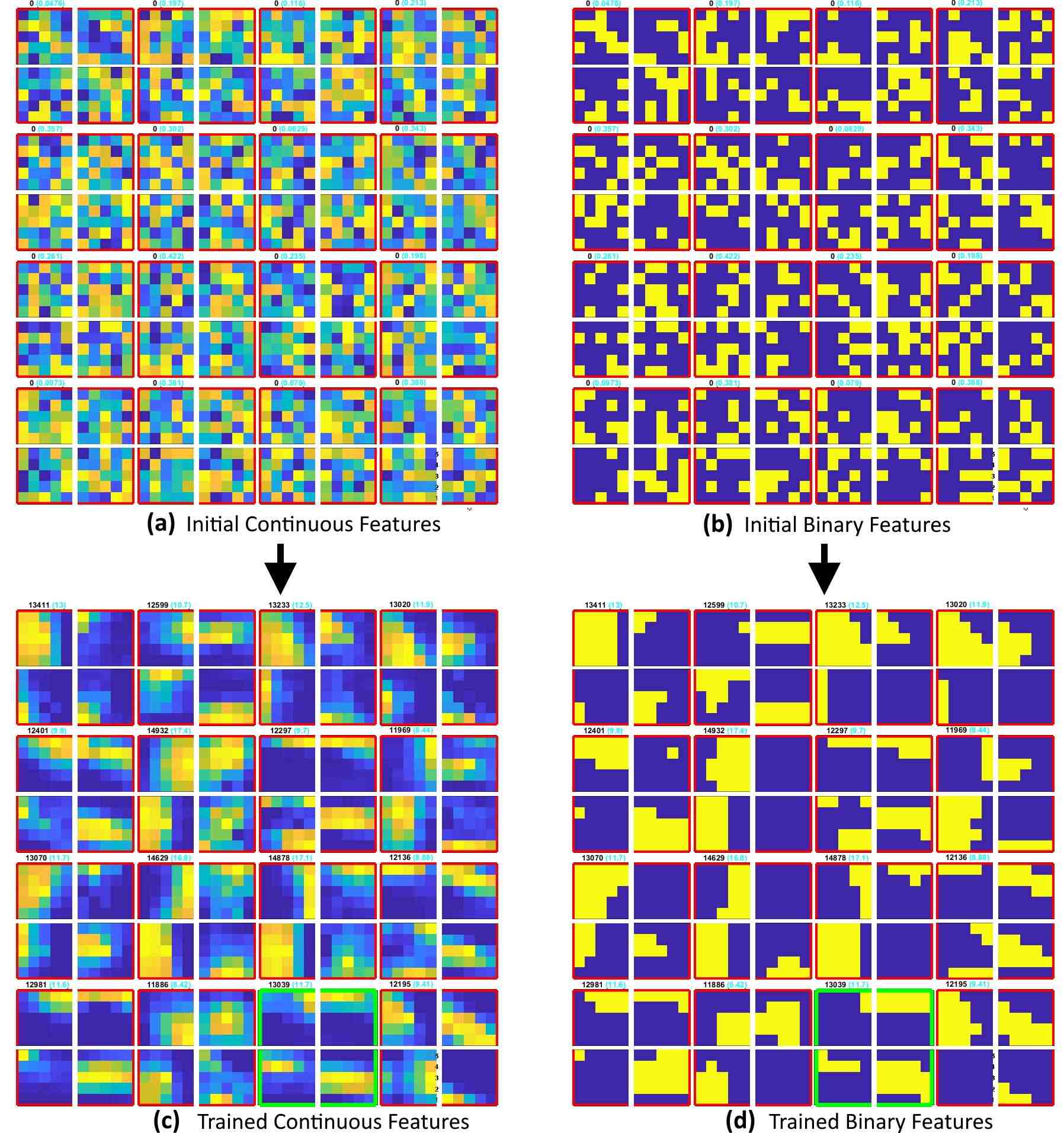}
 \caption{\textbf{Generating 16 binary-valued features on a four-polarity event-based dataset.} (a) Shows the initial random weights of the sixteen four-polarity 5$\times$5 features $\bm w_1$. (b) The binary weighted feature set $\Ddot{\bm w_1}$ with the number of on pixels per feature $m=32$. (c) Final state of the continuous features $\bm w_1$ after training. (d) Final binarized feature set $\Ddot{\bm w_1}$.}
 \label{fig:featureEvol}
\end{figure}

The FEAST algorithm operates via 
three simple rules: When the network receives the $i$th input event $e_i$, the neuron with weight $\bm w_1(n)$ whose cosine distance to the Region of Interest $\bm{ROI}_i$ is smallest, wins the event. This is as long the cosine distance is smaller than the current selection threshold distance. The winning neuron adapts its weights slightly toward the pattern of the $\bm{ROI}_i$. Upon winning the selection threshold of the neuron contracts making the winning neuron more selective around its target spatio-temporal pattern. A second mechanism operates to balance this drive toward greater selectivity whereby every time the entire network of neurons fails to generate an output event in response to an input event, the selection thresholds of all neurons widen making all neurons more receptive. This balancing of the network's selection thresholds and the gradual adaptation of the weights of the network to the observed local spatio-temporal patterns, results in a trained network where all neurons fire at approximately equal rates over the entire training dataset. By balancing network activation the FEAST algorithm ensures that the neurons in the feature extractor network represent the most commonly observed spatio-temporal patterns resulting in a feature set that best represents the underlying training data. Figure \ref{fig:featureEvol} shows an example of the FEAST algorithm training a 16 neuron network on the four-polarity First-AND event-based airplane dataset.

When implementing the FEAST algorithm, the best fitting neuron to an incoming ROI pattern must be determined. This can be achieved most directly via dot product operation which requires $D_1 \times D_1 \times N_1 $ multiplication operations followed by $N_1$ summation operations. However to reduce the hardware resource requirements for this neuron matching operation and to remove the need for hardware implemented multipliers, the continuous-valued feature weights $\bm w_1$ shown in \ref{fig:featureEvol} are converted to binary-valued features. Methods for binarizing images include the Otsus method \cite{otsu1979threshold}, Kittler and Illingworth's minimum error thresholding method \cite{kittler1986minimum} and the Adaptive Binarization method \cite{sauvola2000adaptive}. Here the we use a much simpler equal activation method where for each neuron, the number of 1 valued pixels $m$ is equal. During training, at each presentation, the largest $m$ weights on each neuron are set to one. This method of equal neuron activation allows the unbiased use of AND gates instead of multipliers. When using AND gates as multipliers, if the number of active pixels per feature is not equal, neurons with a lower number of on pixels would activate on more patterns than those with a larger number of on pixels regardless of fitness to the input pattern.

After training, the finalized binary features can be used for inference on the input event stream as an event-based convolutional layer. This results in a first layer feature map $\bm S^1_i$ which can be sampled and processed by a classifier in an event-based manner. In this work, in order to isolate the gains provided by the event-based convolution layer, the input events stream is also converted to an input event surface $\bm S^0_i$ to be sampled and processed by the classifier in an identical manner to the feature map. The generation of the input surface $\bm S^0_i$ and feature layer surface $\bm S^1_i$ are detailed in Algorithm 1 and 2 respectively.

\begin{figure}
 \centering
\includegraphics[width=1\linewidth]{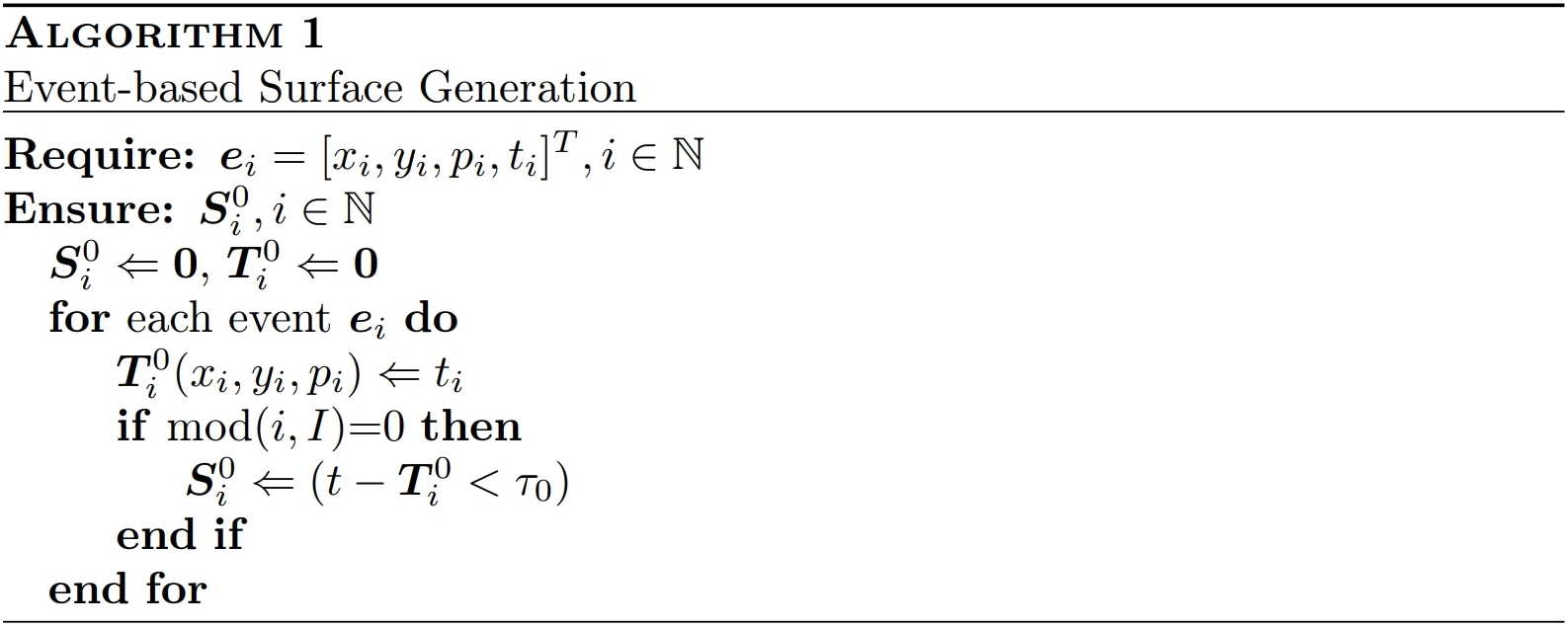}
\end{figure}

\begin{figure}
 \centering
\includegraphics[width=1\linewidth]{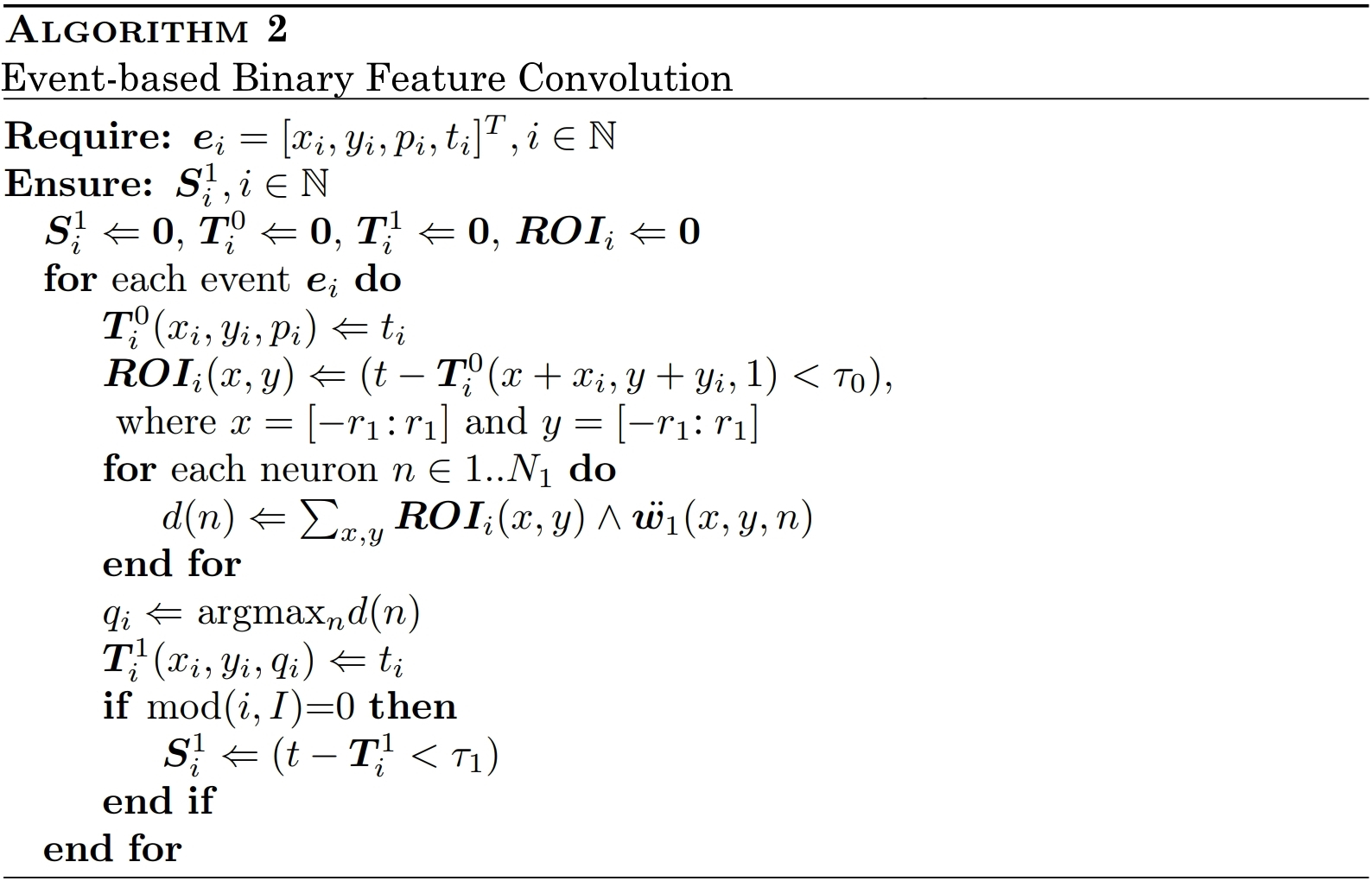}
\end{figure}

\subsection{Alternative Event-based Methods: On-Off Bi-polar and Uni-polar Events}
\label{sec:SPAD/meth/frame2event/diffSpad}
In this section, we introduce two alternative method for comparison to the proposed and implemented First-AND method. In the first method, the difference between consecutive SPAD frames is converted into a sparse On-Off event stream using a simple thresholding operation analogous to those used in other event-based sensors. In the second method, we then augment these On-Off events by introducing uni-polar and bi-polar events. Event-based sensors in general, operate by converting an analog signal (typically pixel illumination in vision) to a sequence of events via a thresholding operation. For the DTOF SPAD imager, this analog signal is the photon time of flight information $Z_k$ which encodes detected depth at the $k$th laser pulse. Algorithm 3 details the generation of On-Off events from the photon time of flight data.

This approach to event generation is more straightforward than the First-AND approach and has the advantage of providing single pixel resolution which is missing in the receptive field based First-AND method. The trade-off, however, is the need for measurement and storage of high-resolution timing data $Z_k$ at each pixel that the First-And approach avoided and which increases the per pixel hardware resource cost.

\begin{figure}
 \centering
\includegraphics[width=1\linewidth]{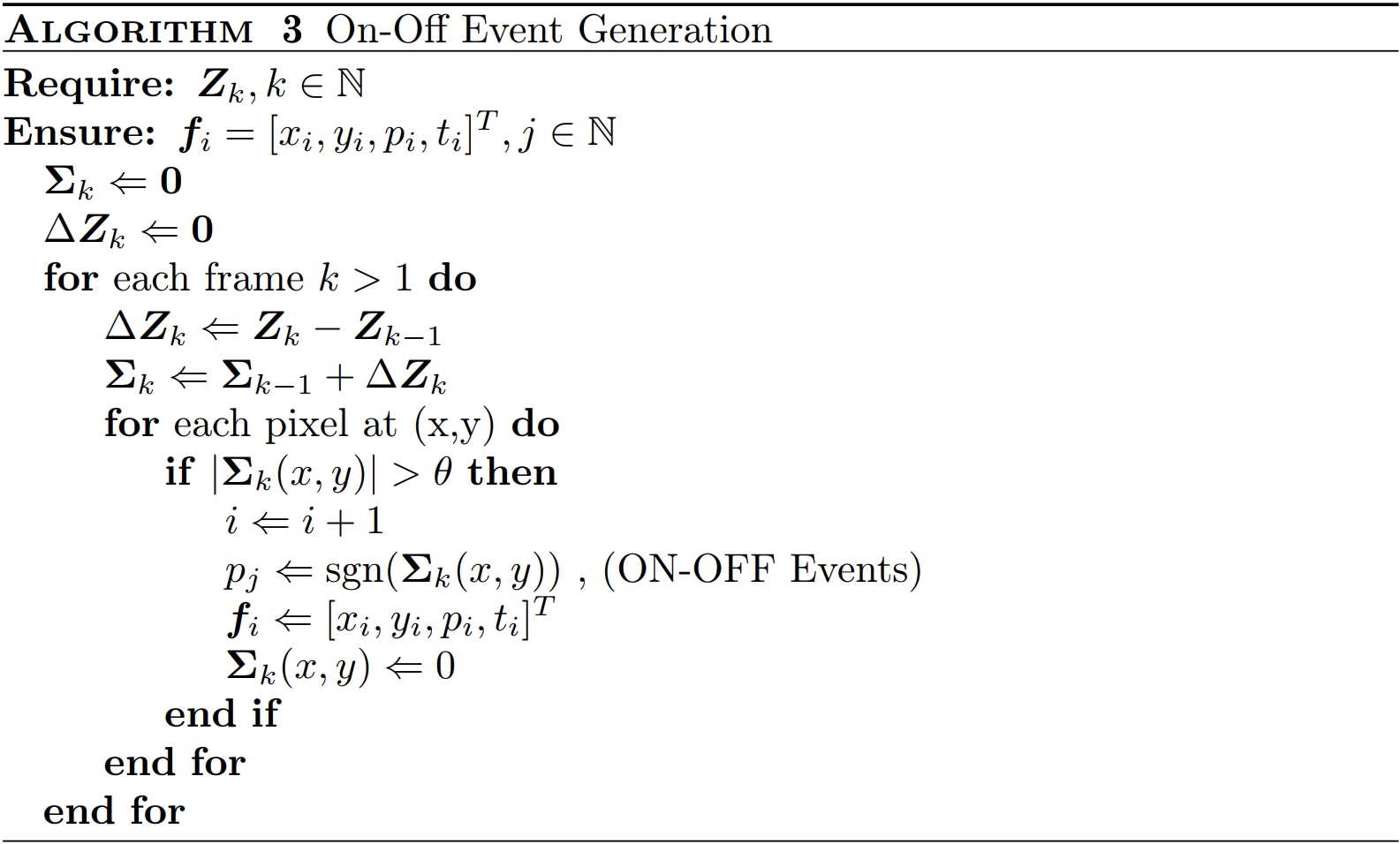}
\end{figure}

Having generated the On-Off events used in standard event-based sensors, we now augment these with two additional event polarities. These two event polarities encode shape invariant local change information via a novel approach. The two event polarities which we name uni-polar and bi-polar events are used to generating a combined On-Off-Bi-polar and Uni-polar (OOBU) event stream.

The bi-polar and uni-polar events are generated using a simple recent event counting operation over the surrounding $3\times3$ pixel region around the current event. As described in Algorithm 4, if the recent events in the 9 pixel region are all On or all Off and their number exceeds a threshold $\phi_1$, then a uni-polar event is generated. If however, both On and Off polarities are present, then if both the On and Off counts are above a threshold $\phi_2$ then a bi-polar event is generated.
In this work the values $\phi_1 = 2$ and $\phi_2 = 1$ were selected through observation of the data.

Since only the local event count is considered, the orientation or structure of recent events does not matter. This results in a unique local feature that is invariant to feature shape, allowing a wide range of different shapes to generate the same event types while still capturing critical local activation information. 

\begin{figure}
 \centering
\includegraphics[width=1\linewidth]{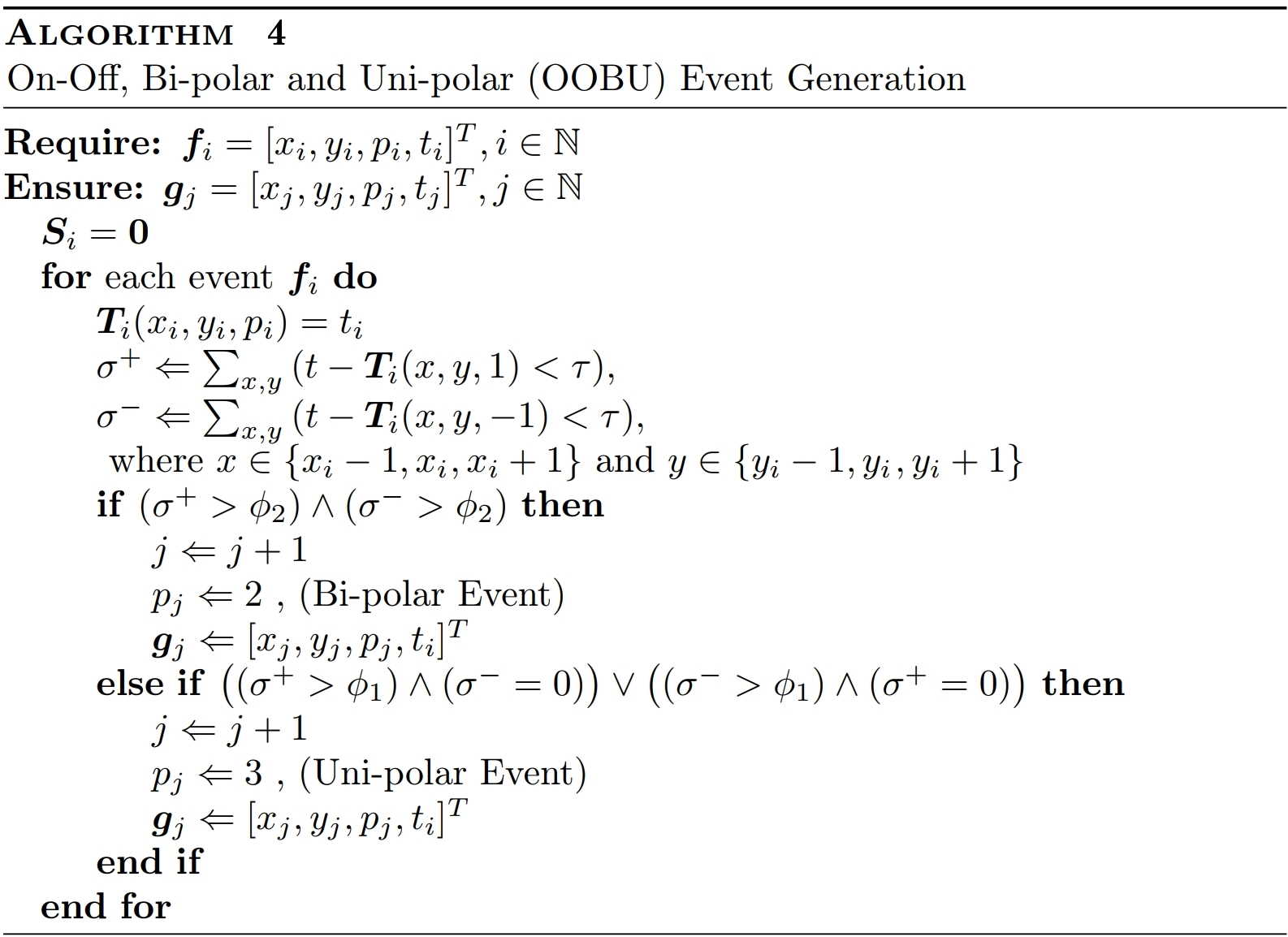}
\end{figure}

\begin{figure}
 \centering
 \includegraphics[width=.5\textwidth]{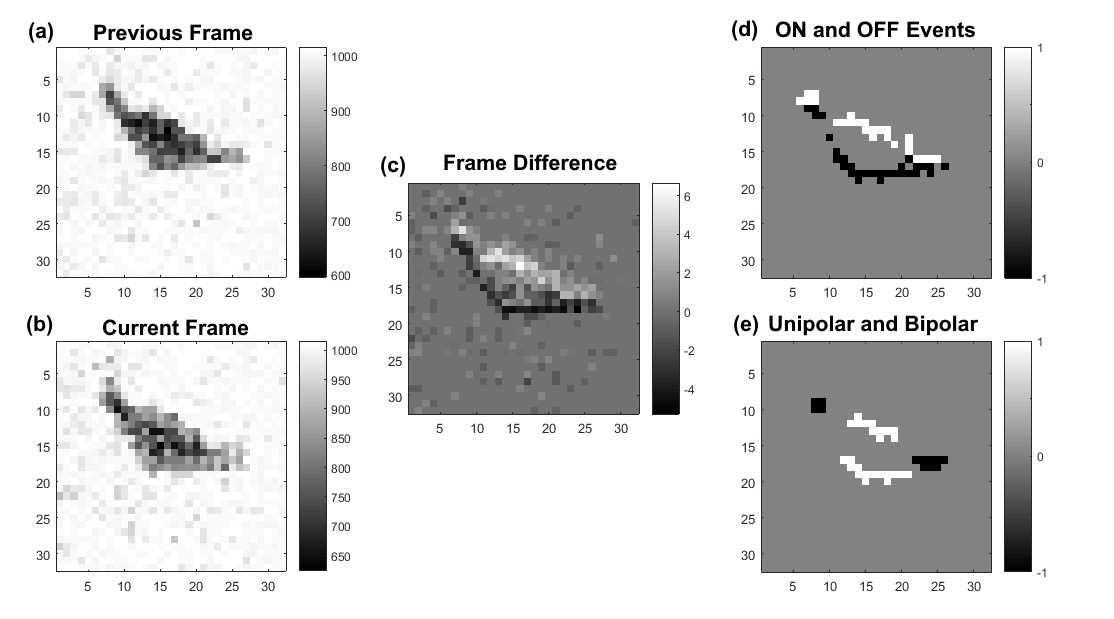}
 \caption{\textbf{Generating On-Off-Bi-polar and uni-polar events from SPAD sensor data.} Panels (a) and (b) show the previous and current captured frames from the SPAD sensor respectively. (c) Shows the frame difference between the current and previous frames. (d) Shows the On and Off events produced via thresholding of the events at +/-$\theta = 2$. (e) Shows the uni-polar (white) and bi-polar (black) events generated via Algorithm 4.}
 \label{fig:diffEvents}
\end{figure}

An example of the On-Off and the uni-polar and bi-polar event streams are shown in Figure~\ref{fig:diffEvents} demonstrating that On-Off event streams faithfully capture the salient spatio-temporal features of the target while the uni-polar and bi-polar events combine local features in a manner that provides distinct, higher scale information to a down-stream processor.

\begin{figure}
 \centering
 \includegraphics[width=.5\textwidth]{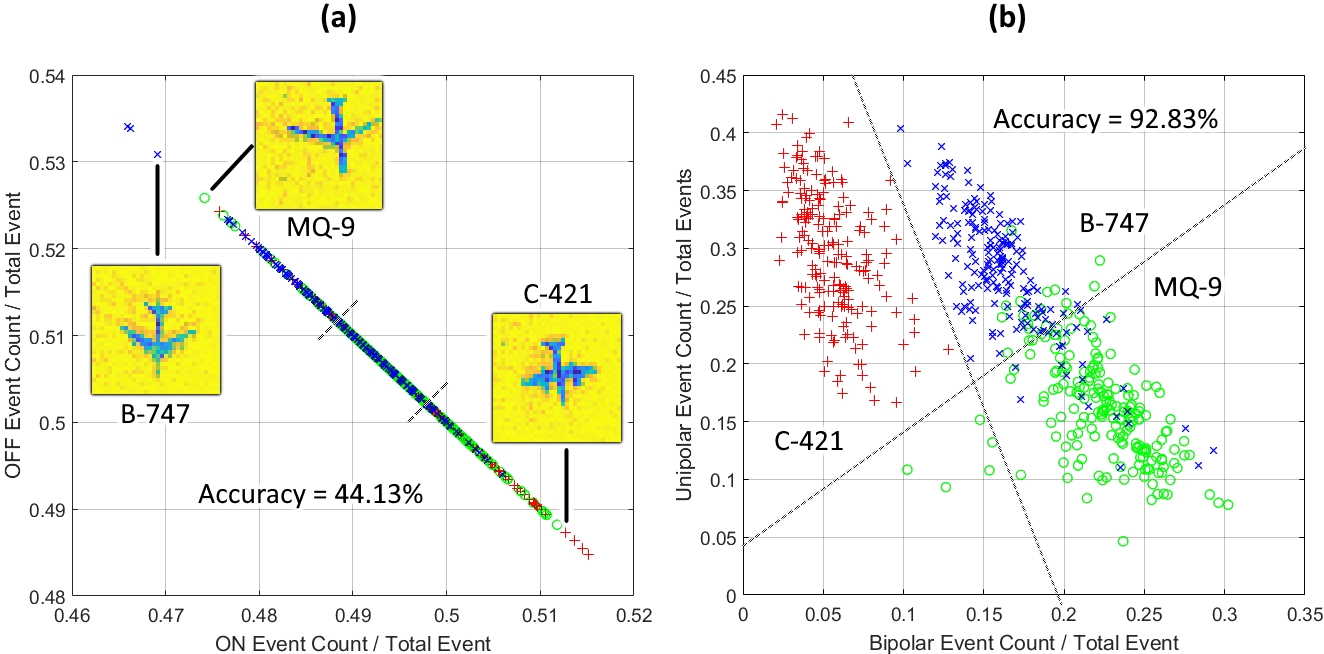}
 \caption{\textbf{Separability of an example three class problem using only event polarity counts.} (a) Plots the ratio of Off event vs On events for each recording for three example airplane classes. The classes are separated using a combination of two thresholds resulting in 44.13\% accuracy. (b) The same test is performed using bi-polar and uni-polar event ratios. Again two thresholds are used to separate the three classes this time resulting in 92.83\% accuracy.}
 \label{fig:diffEventRatios}
\end{figure}

To illustrate the power of uni-polar and bi-polar events in capturing high level salient feature information, a simple three class classification problem is shown in Figure \ref{fig:diffEventRatios}. Here two linear thresholds are combined to separate the three classes using only the event polarity count information. Figure \ref{fig:diffEventRatios}(a) shows that simple examination of the On and Off event counts is not a useful method of discriminating the three example classes. For this example the best accuracy achievable using two separating On-Off event count ratio thresholds is 44.13\% accuracy which is only slightly above chance 33.33\%. In contrast, as shown in Figure \ref{fig:diffEventRatios}(b), when bi-polar and uni-polar events are generated from the On-Off event stream, a simple bi-polar uni-polar event count test results in 92.83\% accuracy when two event count ratio thresholds are used in combination. While this extremely simple event counting method does not extend to more challenging tasks (such as the full 15 class SPAD dataset) this simple example illustrates the significant discriminatory power of OOBU events. In this work, we show that when OOBU events streams are processed in a more sophisticated manner, they outperform other types of event streams and produce the highest performing results of the dataset.

\begin{figure}
 \centering
 \includegraphics[width=0.5\textwidth]{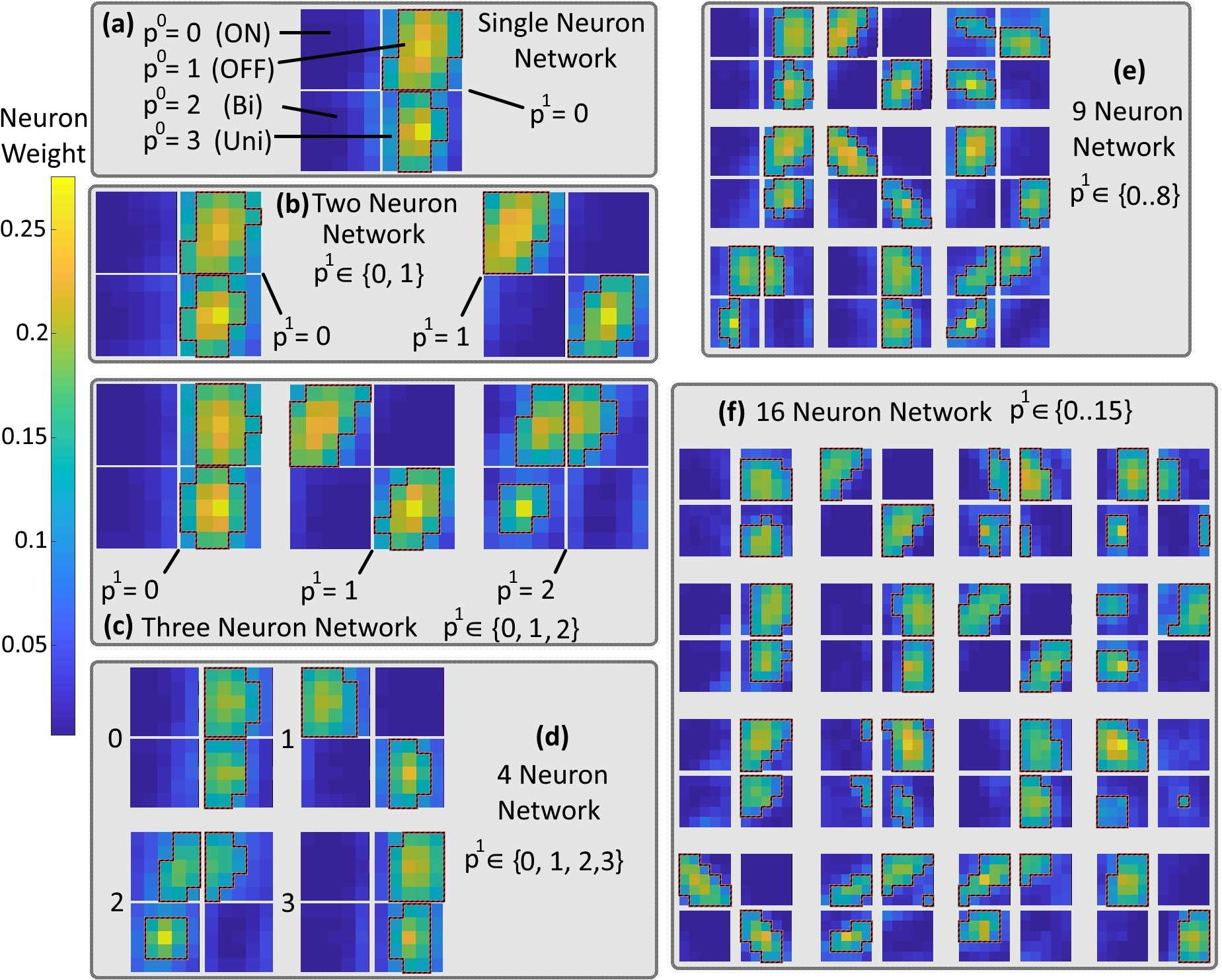}
 \caption{\textbf{Trained OOBU neuron weights across network sizes.} (a) Shows a trivial single neuron network. The four images represent the On ($p^0=0$), Off ($p^0=1$), bi-polar ($p^0=2$) and uni-polar ($p^0=3$) feature weights of the single neuron ($p^1=1$). This single neuron network simply serves to show the dominant spatio-temporal pattern present in the dataset which is only Off events ($p^0=1$) occurring together ($p^0=3$). (b) A two neuron network ($p^2={0,1}$) shows the next most dominant observed pattern is the On events ($p^0=1$) occurring together ($p^0=3$). Note that the first neuron in this network is nearly identical to that in (a). (c) The third most dominant pattern is the On and Off events occurring together in a diagonal pattern. (d) The fourth most dominant pattern is a second variant of the OFF event occurring alone on the right-hand side of the ROI. Note that the presence of this variant results in the first neuron being trained in a complementary manner with the Off events occurring on the left-hand side of ROI such that the summation of neuron 1 and 4 would approximately equal the single dominant neuron in (a). Panels (e) and (f) show networks of 9 and 16 neurons with increasingly complex features.}
 \label{fig:neuronWeights}
\end{figure}

We now combine the OOBU events generated from the frame-based SPAD dataset into a single event stream and process it through the same surface generation and feature extraction algorithm described in Algorithm 1 and Algorithm 2 using identical training architecture and learning parameters. In doing so we are able to combine the information from the On-Off, uni-polar and bi-polar event streams into a single feature extractor network. Figure \ref{fig:neuronWeights} shows the spatio-temporal patterns extracted at each network size from the SPAD OOBU event-based data stream. Figure \ref{fig:neuronWeights} demonstrates how the FEAST algorithm extracts the dominant patterns in the dataset for any given network size and how the information contained in the On, Off uni-polar and bi-polar event streams can be combined to provide powerful discriminatory features.

\subsection{Pooling, Surface Sampling and Classification}
\label{sec:SPAD/meth/poolLinClass}
After the feature extraction operation performed by FEAST is complete, the target region of size $A_x \times A_y$ on the time surface is selected via a surface summation and thresholding operation described in \cite{afshar2018investigation} and implemented for real-time GPU based platforms in \cite{mau2019}.

After the $A_x \times A_y$ target region is selected, the variable sized 2D area from the surface must be mapped to the statically sized classifier input layer. To perform this mapping we explore two alternative methods.
In the first method, which we call 1D pooling, the $A_x \times A_y$ region is summed across rows and columns resulting in two one-dimensional vectors $V_x$ of size $A_x \times 1$ and $V_y$ of size $A_y \times 1$. These two vectors are then re-sampled to two $L \times 1$ vectors. To speed up and simplify these re-sampling operations, the input data was first cropped or zero buffered and resampled using a zero-order-hold operation which was implemented via a pre-calculated Look Up Table. This hardware optimized method was introduced in \cite{mau2019}. 
In the second method, which we call 2D pooling, the $A_x \times A_y$ target region is re-sized to a two-dimensional image of size $L \times L$. For this method a 2D resample function using linear interpolation between adjacent values was used.

The 1D pooling method significantly simplifies the implementation of the resampling and also provides the added benefit of a classifier with a smaller input layer. However the 2D pooling method captures significantly more information and as we show in this work, results in higher recognition performance in most cases, creating design trade-offs that require investigation. In a similar way, the size of the classifier input layer which is $m = L \times 2$ in 1D case, and $m = L \times L$ in the 2D case, can also affect the accuracy. For this reason in this work we perform all trials over a range of pool sizes ($L \in \{1, 2, 3, 4, 6, 8, 12, 16, 24 \}$) to investigate these effects and provide useful design guidelines for hardware implementation.

Another important hyper-parameter in the operation of the feature extraction and classification system is the frequency of surface or image sampling for the processing by the classifier.

For the frame-based data, the time interval between classification operations was selected as 80 microseconds or every 8th laser pulse. This time interval resulted in a total of 1.22 million classification operations over the entire dataset or approximately $50.51 +/-8.07$ classification operations per recording. This total number of operations was then used to normalize the number of classification operations on the event-based dataset to provide an approximately equal number of input samples to the classifier enabling an unbiased comparison of the frame-based and event-based systems. For the First-AND event streams, keeping the total number of classification operations constant results in an inter classification event interval of 51 events. For the On-Off and OOBU events, the interval between classification operations becomes 74 and 201 events respectively. By operating the classifier in this event-based manner, the rate of processing becomes dependent on the level of salient change in the field of view as opposed to constant in the frame-based approach.

All classification tests in this work were performed on the full 15 airplane, 24000 recording augmented dataset. The dataset was split randomly into a 21600 recording (90\%) training set and 2400 (10\%) test set. All tests were repeated over $n$ = 20 trials using randomized splits of the dataset recordings. The original frame-based dataset was converted to the equivalent event-based datasets via the methods described in Section \ref{sec:SPAD/meth}.
All tests were performed using a simple linear classifier to probe the performance of each configuration in an unbiased manner and without the introduction of additional computational complexity. The input to the linear classifier consists of a resized target region from a sampled time surface ($\bm S^0_i$ or $\bm S^1_i$) which is vectorized into a $1\times m$ input vector $\bm u$. where, $m = L^2$ for the 2D pooling case and $m = 2L$ for the 1D pooling case. The output of the linear classifier is a $1\times n$ predicted output vector $\bm \hat{v}$ with $n$ being the number of output classes which in this case is 15. The predicted output vector $\bm \hat{v}$ is calculated via:

\begin{equation}
 \label{eq:outputEqualsWeightTimesInput}
\bm v = \bm W \bm \hat{u}
\end{equation}
where $\bm W$ is the trained $m\times n$ weight matrix.

To determine the winning output class $j$ during inference, an argmax operation is performed on the predicted output vector $\bm v_i$:

\begin{equation}
 \label{eq:winningLinClass}
j = \text{argmax}_{i \in [1..n]}(\bm \hat{v}_i)
\end{equation}

For the process of training and testing, all sampled input vectors $u$ are concatenated to form the $o\times m$ input matrix $\bm U$ where $o$ is the number of samples over the entire dataset. A corresponding $o\times n$ ground truth output matrix $\bm V$ is also generated using a 'one-hot coding' scheme with the actual class for each sample set to 1 and all others set to zero.

The input matrix is then split into the 21600 recording training input matrix, $\bm U_a$ and 2400 recording testing input matrix $\bm U_b$. The corresponding output matrix $\bm V$ is similarly split into $\bm V_a$ and $\bm V_b$. The classifier is trained via a calculation of the pseudoinverse solution to the weights mapping the input activation layer or feature layer to the output classes as given by: 

\begin{equation}
 \label{eq:calculateLinWeights}
\bm W = (\bm U_a^T \bm V_a)^T (\bm U_a^T \bm U_a + \lambda \bm I)
\end{equation}

Where $\bm I$ is the identity matrix and $\lambda = 0.1$ is the regularization factor used for applying ridge regression \cite{tapson2013learning}. In cases where the pseudoinverse operation could not be performed in a single pass (due to memory constraints) the equivalent online method was used \cite{VanSchaik2015b}. In either approach, the training operation is deterministic and repeatable such that the source of variance in classification accuracy is exclusively due to the feature extraction operation and the random splitting of the dataset.

\subsection{Event-based Processor Implementation on FPGA}
\label{sec:SPAD/meth/FPGA}
The event-based binary feature convolution and classifier system was implemented in FPGA hardware on a Cyclone IV E platform. By implementing the system on a relatively small FPGA platform, we demonstrate the utility of the proposed system in applications where hardware resources are limited.

\begin{figure}
 \centering
 \includegraphics[width=.35\textwidth]{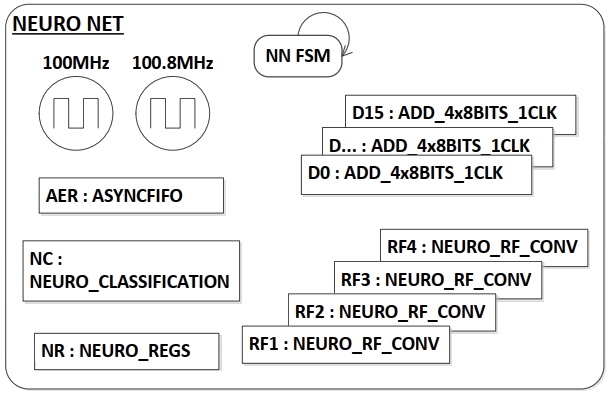}
 \caption{\textbf{Neuro Net Entity Clock Domains and Instantiated Components.}}
 \label{fig:fpgaComponent}
\end{figure}

The input to the system is defined as the four-polarity event streams (here referred to as the 4 RFs for Receptive Fields). 
The top-level subsystem called NEURO\_NET, incorporates a $5\times5$ ROI patch size with 16 feature neurons. The asynchronous FIFO component shown in Figure \ref{fig:fpgaComponent}, contains the encoded AER events received via the event-based SPAD sensor for processing. The FIFO holds $2^9$ words (512) of 32-bits in length. The FIFO is accessed via an FSM (Finite State Machine) which sequences the reading and processing of events and is used to control access to the DDR2 controller and synchronization logic between the 133 MHz clock domain of the DDR2 Controller and the clock domain in which the access originates. The NEURO\_REGS component contains the local registers that hold the binary features $\Ddot{\bm w^1}$ used by the NEURO\_RF\_CONV components. The registers are programmed via software during initialization. The NEURO\_RF\_CONV components also contain the binary convolution logic of the feature set and the ROI patch loaded from the recent time surface $\bm S^0$. Four NEURO\_RF\_CONV components are instantiated, one for each event polarity of the event stream. The output of this component is the convolved feature map. The ADD\_4x8CITS\_1CLK components contain the logic for the addition of the convolved features maps of each of the 4 polarities and the 16 features. The NEURO\_CLASSIFICATION component contains the local registers that hold the classifier weights $W$ which are pre-calculated offline and programed via software during initialization, the logic to determine the winning neuron from the sum of the convolved feature maps and the logic to perform the classification.

\begin{figure}
 \centering
 \includegraphics[width=.5\textwidth]{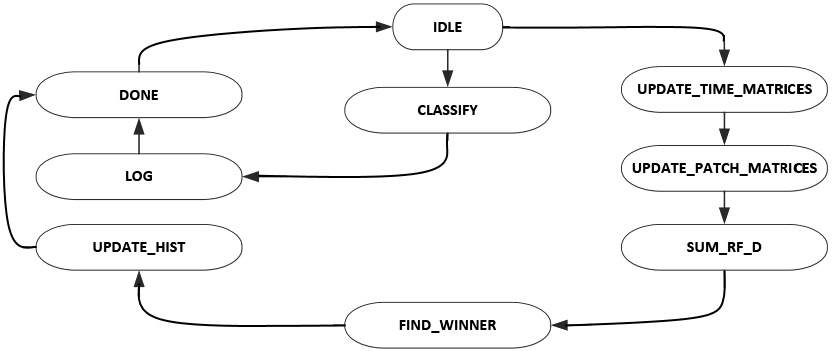}
 \caption{\textbf{NEURO\_NET Finite State Machine (FSM).}}
 \label{fig:fpgaFsm}
\end{figure}

Figure \ref{fig:fpgaFsm} displays the FSM for the NEURO\_NET component. In state 'IDLE' the AER FIFO is interrogated and if an AER event exists the event is read and state transition to 'UPDATE\_TIME\_MATRICES' occurs. In state 'UPDATE\_TIME\_MATRICES' the AER event is decoded to the row/column addresses along with the neuron triggering the event and the event time. The corresponding time address in the DDR SDRAM is updated with the event time and state transition to 'UPDATE\_PATCH\_MATRICES' occurs. In state 'UPDATE\_PATCH\_MATRICES' the ROI patch is read from the DDR2 SDRAM starting with the first polarity. With an ROI patch size of $5\times5$ pixels and 4 polarities, a total of 100 DDR2 SDRAM addresses are read. As each pixel is read, the corresponding NEURO\_RF\_CONV component is signaled and the read pixel patch time convolution occurs. Once all patches have been read and processed, the state transition to 'SUM\_RF\_D' occurs. In state 'SUM\_RF\_D' the outputs of the 4 NEURO\_RF\_CONV components is summed and state transition to 'FIND\_WINNER' occurs. In state 'FIND\_WINNER' the NEURO\_CLASSIFICATION component iterates through the summed convolution components to determine the max count and determine the winner neuron. State transition to 'UPDATE\_HIST' then occurs and in this state the feature output or histogram is updated from the winning neuron(s). The state then transitions to 'IDLE' through state 'DONE'. In state 'IDLE' if the classification time has expired and a prescribed number of events have been processed, state transition to 'CLASSIFY' occurs. In state 'CLASSIFY' the NEURO\_RF\_CONV component performs the vector dot product operations on the feature maps. The result of this state is the classification and state transition to 'LOG'. The state 'LOG' samples the current feature maps for each neuron along with the resultant dot product output values which are used for classification. The log is added to a FIFO and can be retrieved by the software for post-processing. A state transition to 'IDLE' then occurs through state 'DONE'. This process repeats itself so long as AER events are available.

\subsection{Comparison to Frame-based System}
\label{sec:SPAD/meth/GPU}
To provide a comparison for the performance of the event-based feature extraction networks, equivalent frame-based systems with identical architectures and training methodologies were developed and tested. Here the event-based feature extractors are replaced with convolution and max pooling operations with the same feature sizes. In this way the frame-based networks precisely replicate the event-based operations with the only difference between the two methods being the extra, arguably unnecessary, convolution operations performed in the frame-based system on the parts of the image exhibiting no significant change i.e. those with no events. Following the convolutional layer, the same pooling methods and linear classification operations were performed for all tests providing an unbiased comparison between the frame-based and event-based systems.

A subset of these frame-based feature extraction networks were implemented on an embedded NVIDIA Jetson TX2 board \cite{mau2019}. This hardware implementation aimed to demonstrate the feasibility of realizing a high-speed hardware effiecent feature extraction and classification system for noisy low-resolution SPAD imagers. To further simplify the implementation the simpler 1D pooling method was used in this work. In the next section we compare the classification performance results of a range of frame-based systems to equivalent event-based systems.

\section{Results}
\label{sec:SPAD/res}

\subsection{Data Generation Rates}
\label{sec:SPAD/res/datarates}

\begin{figure}
 \centering
 \includegraphics[width=.4\textwidth]{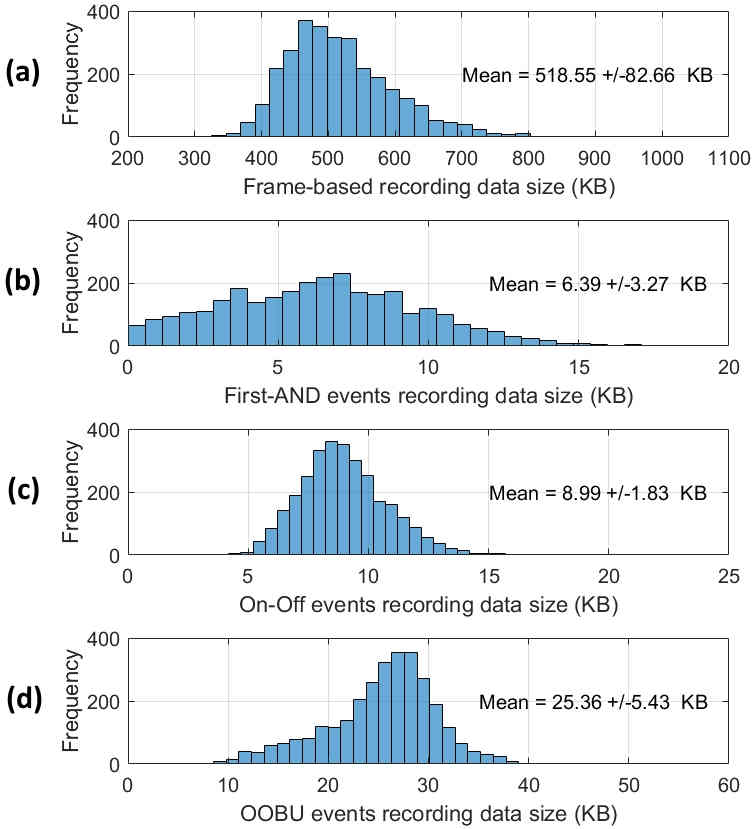}
 \caption{\textbf{Reduction in data size by event-based conversion.} (a) shows the distribution of size of the frame-based SPAD recordings. Panels (b), (c) and (d) show the size distributions of the First-AND, On-Off and OOBU event streams respectively.}
 \label{fig:dataRates}
\end{figure}

The recorded frame-based SPAD imaging dataset was converted to a First-AND event-based data stream via simulation of the implemented First-AND circuit described in Section \ref{sec:SPAD/meth/frame2event/AsicSpad}. In addition, the dataset was processed using Algorithms 1 and 2 to generate the On-Off and OOBU event-based data streams.
As shown in Figure \ref{fig:dataRates}, the conversion of frame-based SPAD data to an event stream significantly reduces the recording size and thus the data-rate of the processor with associated savings in processing power and improved response time. The First-AND conversion method results in an 81 fold reduction in data-rate whereas the On-Off and OOBU methods result in 57 and 25 fold reductions respectively. Having examined the data-rates generated from the different methods, we now compare the classification performance of the First-AND, On-Off and OOBU event streams to the original frame-based SPAD imaging dataset.

\subsection{Classification}
\label{sec:SPAD/res/class}

 The tests presented in this section cover the raw frame-based dataset as well as the event-based methods described. We further test the effect of different pooling methods as well as processing by random and trained feature extraction networks. The tests also examine in detail the effect of pooling window sizes and network sizes on performance. All test were performed using a linear classifier mapping frames-based images and event-based samples of time surfaces to the output classes in a repeatable manner. Where applicable, fixed-point precision was used for simulation parameters mirroring the fixed-point FPGA implementation described in Section \ref{sec:SPAD/meth/FPGA} and the ASIC implementation in Section \ref{sec:SPAD/meth/frame2event/AsicSpad}.

\begin{figure}
 \centering
 \includegraphics[width=0.5\textwidth]{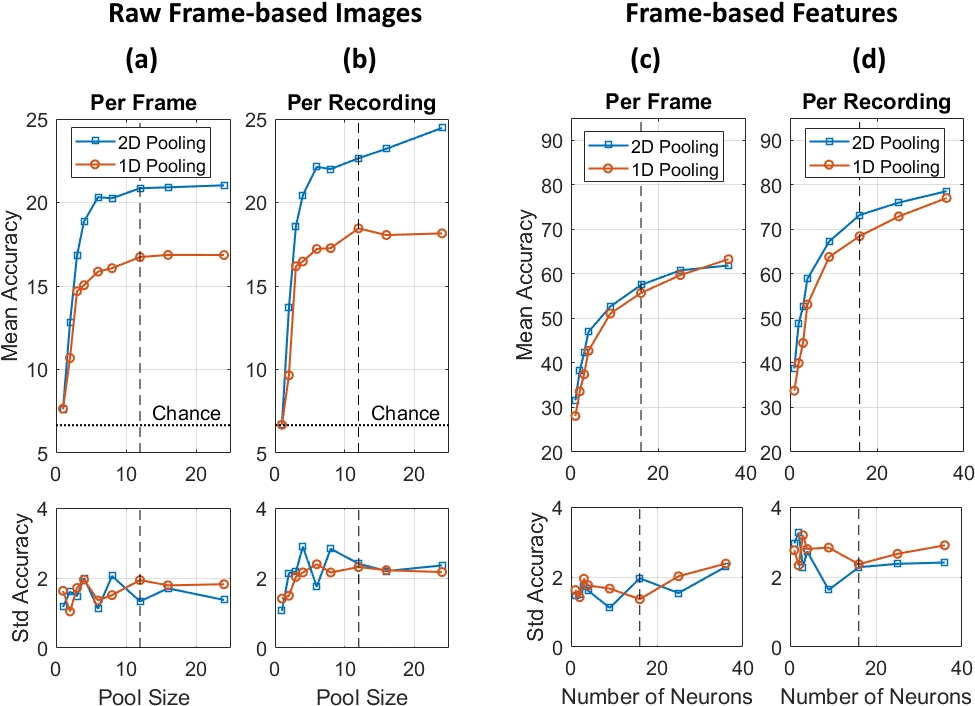}
 \caption{\textbf{Classification accuracy on the frame-based SPAD airplane dataset across a range of parameters.} Panels (a) and (b) show per frame and per recording accuracy respectively where classification is performed directly on the raw SPAD images via a linear classifier. The top and bottom two panels show the mean and standard deviation of classification accuracy respectively. Results for both two-dimensional pooling and one-dimensional pooling are plotted as a function of pool size and compared to chance 1/15 = 6.7\%. The vertical dashed line at $L = 12$ indicates the pooling window size chosen in the subsequent tests. Panels (c) and (d) show per frame and per recording accuracy respectively after feature extraction as a function of the number of feature extracting neurons. The dashed vertical line at $N = 16$ marks the network size chosen for FPGA implementation. All results presented are over $n$ = 20 independent trials.}
 \label{fig:rawFrameBasedResults}
\end{figure}
Figure \ref{fig:rawFrameBasedResults} shows the classification results when using frame-based processing on the SPAD dataset. The results are organized in per frame and per recording accuracy results. For the per recording accuracy measure, the class with the highest number of winning frames is selected as the correct class. Unsurprisingly, since this process effectively performs a pooling and max operation over the information in all the frames of a recording, the per recording classification accuracy is consistently above that of the per frame accuracy measure.

It is clear from the accuracy results in Figure \ref{fig:rawFrameBasedResults} that the 2D pooling almost always outperforms 1D pooling. However the 1D pooling method is significantly simpler to implement in hardware and faster to compute in software motivating its investigation and comparison to the 2D pooling method. Given hardware constraints, these comparisons provide valuable information on the resource versus performance trade-offs which are critical during the hardware design stage.

The first point in Figure \ref{fig:rawFrameBasedResults}(a) at $L = 1$, collapses all information in each frame to a single number. As expected, this global pooling of the entire raw image produces an identical accuracy for both the 1D and 2D pooling methods that is close to chance. This result effectively demonstrates that, as expected, the mean value across the pixels of the image provides approximately zero information about the target class. As the size of the pooling window $L$ increases, the classification accuracy rises sharply before stabilizing above $L = 12$ pixels since little additional information can be generated by increasing the pooling window resolution to or above the original $A_x \times A_y$. Thus the best results achievable using the raw frame-based SPAD data, a pooling layer and a linear classifier is accuracy that is below 25\%. The per recording accuracy measure shown in \ref{fig:rawFrameBasedResults}(b) is similarly poor providing only slightly higher accuracy at the larger pool sizes.

Figure \ref{fig:rawFrameBasedResults}(c) and (d) show the classification accuracy of trained frame-based feature extraction networks with identical architecture and training parameters as the event-based networks discussed in Section \ref{sec:SPAD/meth/frame2event/feast}.
The first point in Figure \ref{fig:rawFrameBasedResults}(c), at $N=1$, $L=12$ and accuracy of approximately 30\%, represents a trivial convolution of the SPAD frames by the single commonest feature in the dataset. Capturing slightly more spatial information than the raw image, this trivial solution performs only slightly better than the raw frame results of (a) with pool size $L=12$. Here, the additional information derived from the incorporation of local spatial information in the convolution operation provides approximately 10\% improvement in accuracy. As the number of feature extractors is increased, the classification accuracy increases to slightly above 60\% and below 80\% for the per frame and per recording measures respectively at $N=36$ neurons. While every increase in network size improves system accuracy, there are diminishing returns with each layer size increase, a pattern that we see consistently in all tests. Note that the comparison of the 1D pooling and 2D pooling methods shown in Figure \ref{fig:rawFrameBasedResults}(c) and (d) demonstrate only a slight advantage in favor of the 2D pooling method thus validating the hardware design choice of implementing the simpler 1D pooling method in \cite{mau2019}.

\begin{figure}
 \centering
 \includegraphics[width=.5\textwidth]{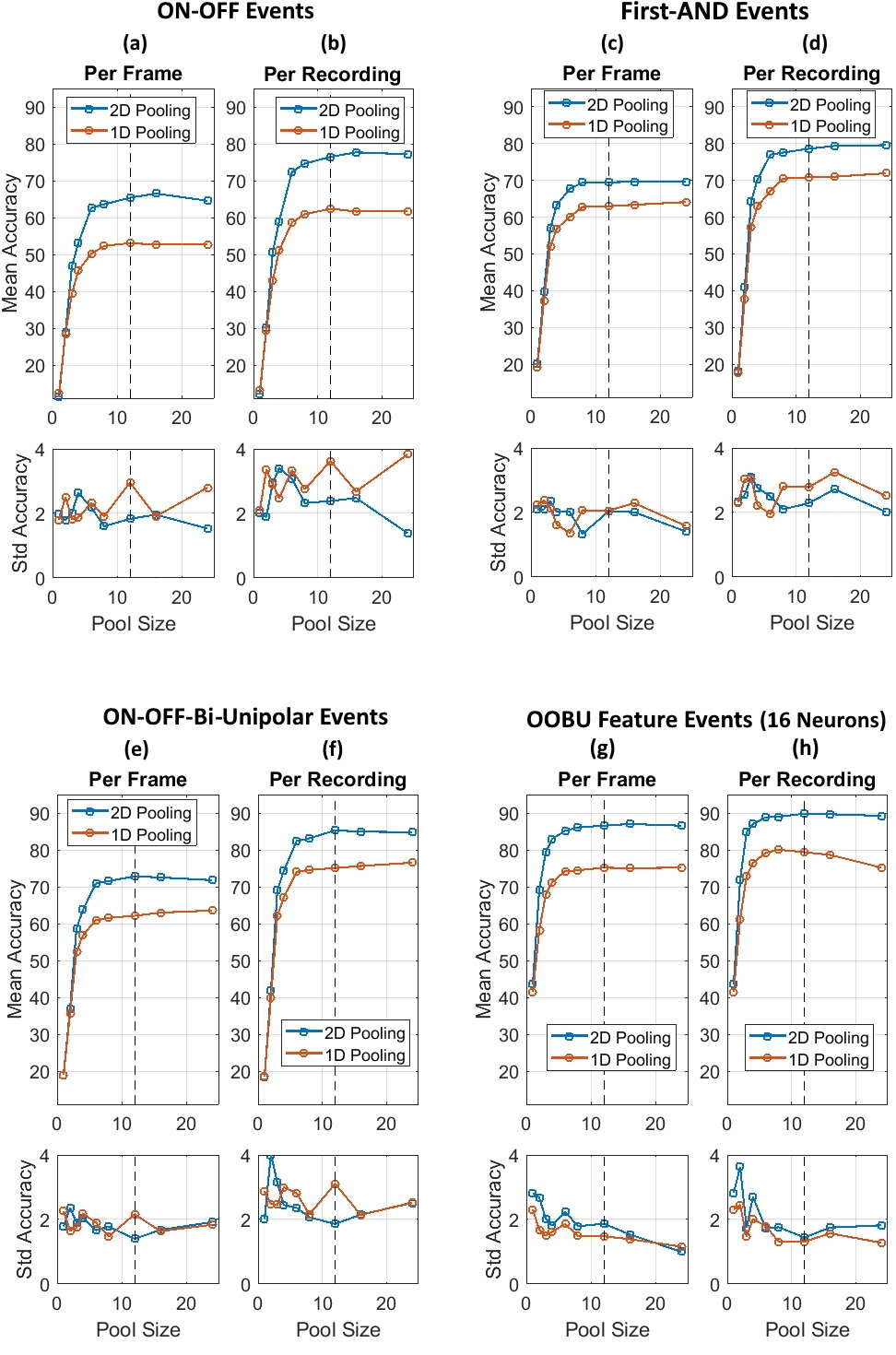}
 \caption{\textbf{Classification accuracy on event-based data streams generated from the SPAD the dataset.} Panels (a) and (b) show the per frame and per recording accuracy results of the On-Off event streams. Panel (c) and (d) show the same for First-And events, (e) and (f) for OOBU events and (g) and (h) show the same for feature events generated from OOBU events using 16 FEAST neurons. All results are shown as a function of the pool window size $L$. The vertical dashed line indicates the chosen window size $L=12$ used in subsequent tests.}
 \label{fig:poolClassificationResults}
\end{figure}
Figure \ref{fig:poolClassificationResults} shows the classification performance of the proposed event generation methods. The classification results show a large increase in accuracy for all event-based methods relative to the original frame-based SPAD dataset shown in Figure \ref{fig:rawFrameBasedResults}(a). The event stream with the lowest accuracy is that of the On-Off events shown in Figure \ref{fig:poolClassificationResults}(a). The first point plotted is at pool size $L=1$ which shows per accuracy slightly above 10\%. This is equivalent to only using the event polarity count for classification which unsurprisingly results in the lowest accuracy. As the pool size is increased above $L>8$ the per frame accuracy rises reaching approximately 55\% and 65\% for the 1D and 2D pooling methods respectively. The results in (b) follow a similar pattern with per recording accuracies reaching 62\% and 76\% respectively for the 1D and 2D pooling methods. The First-AND event results are shown in panels (c) and (d). Here the 2D pooling accuracies are slightly above those of On-Off events. The relative improvement in accuracy is even greater for the 1D pooling method again motivating its use in hardware. The OOBU event accuracy results shown in (e) and (f) are consistently higher than those of both the On-Off and First-AND events making OOBU events the clear winner of the three approaches.

Finally panels (g) and (h) of Figure \ref{fig:poolClassificationResults} show the accuracy achieved via the addition of an event-based feature extraction layer. This configuration is used in the hardware implemented system described in \ref{sec:SPAD/meth/FPGA} with OOBU events serving as inputs to 16 trained features. The performance of the full system is examined as a function of different pooling methods and pool sizes. The per frame and per recording accuracies of the feature extraction layer are the best of all the event streams tested, starting from slightly above 40\% accuracy and reaching 75\% and 87\% per frame accuracy for the 1D and 2D pooling methods respectively. The per recording accuracies are even higher at 79\% and 90\% for the 1D and 2D pooling methods respectively. Here the $L=1$ result at above 40\% accuracy and the $L=12$ result at 90\% accuracy are both remarkable given the complexity of the 15 class view-invariant airplane classification dataset, the simplicity of the applied methods and the significantly lower performance on the frame-based dataset using identically structured and trained classification architectures. These accuracy results together with the clear data-rate advantages detailed in Section \ref{sec:SPAD/res/datarates}, highlight the suitability of the use of event-based compressive sensing approach to the high noise SPAD time of flight imaging data.

\begin{figure}
 \centering
 \includegraphics[width=.5\textwidth]{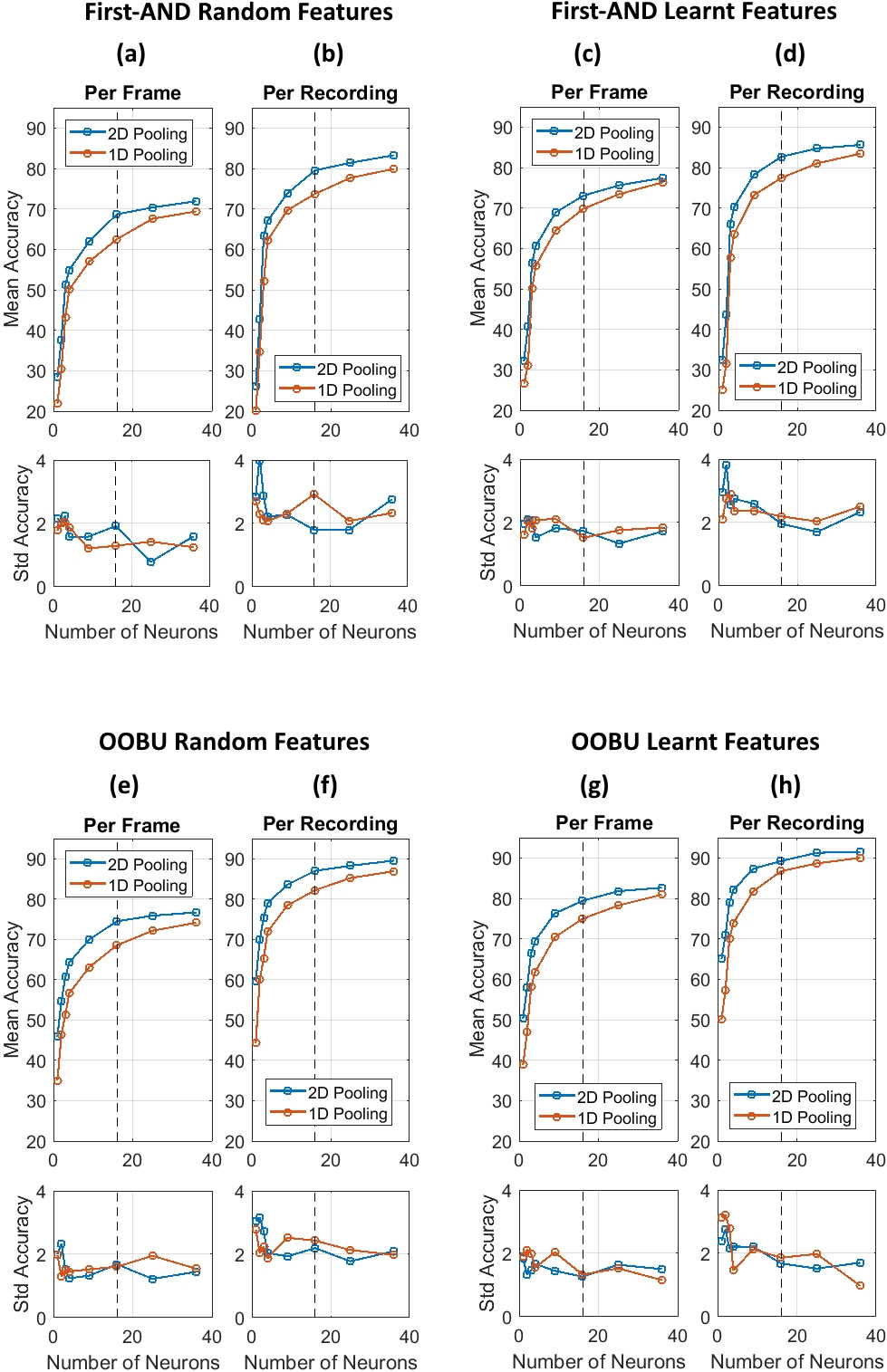}
 \caption{\textbf{Classification accuracy using feature event streams generated from random binary networks and trained binary networks operating on the proposed First-AND and OOBU event streams} (a) and (b) show per frame and per recording classification accuracy using First-AND events processed through a random binary feature extraction layer of varying size and (c) and (d) show the same with the random binary features replaced by trained binary features. (e) and (f) show per frame and per recording accuracy results of random binary features operating on OOBU events and (g) and (h) show the same for trained binary features. The dasehd vertical line shows the $N = 16$ configuration selected for implementation of the event-based FPGA processor described in Section \ref{sec:SPAD/meth/FPGA}.}
 \label{fig:NeuronClassificationResults}
\end{figure}

Having investigated the effects of different pooling methods on the various event-based data streams, we now investigate the effect of the feature extractor network size on classification accuracy. Since the size of the feature extractor network affects hardware resource consumption and/or processing speed, we seek to determine the smallest network which provides an acceptable level of performance given the resource constraints and speed requirements. Figure \ref{fig:NeuronClassificationResults} shows accuracy results for the First-AND and OOBU event streams which provided the highest performance in the pooling test experiments. Figure \ref{fig:NeuronClassificationResults}(a) and (b) shows the per frame and per recording accuracies of the First-AND event stream processed by random binary-valued feature extraction networks. By using random features with identical network architectures and processing methods, the improvements gained via feature training alone can be isolated. The first points on panels (a) and (b) represent trivial single neuron feature extractors. As the layer size increases the mean classification accuracy increases rapidly reaching slightly above 70\% at the highest network size tested which is $N=36$. These random feature results provide a baseline for evaluating the trained binary networks whose results are shown in panels (c) and (d). The results show that trained features consistently outperform random features. Here as in the raw event stream tests, 2D pooling still consistently outperforms 1D pooling, here however, the margin is smaller. This result is expected since the feature extraction operation projects the raw event stream onto a large number of sparsely populated feature surfaces such that when the surfaces are pooled via the 1D method, less information is lost in comparison to the 2D pooling method. In other words, as the size of the feature extraction layer expands, the effect of information loss due to 1D pooling becomes less significant. 

In panels (e) to (h) the same comparison between random features and trained features is performed, this time on the OOBU events. We again see that as with the First-AND case, trained OOBU features outperform random ones. And again we see that OOBU events consistently outperform First-AND events, this time when processed through an event-based feature extraction network. The results in panel (g) and (h) represent the highest accuracy achieved on the dataset where at a layer size of $N=36$, a per frame accuracy of 82.64\% and a per recording accuracy 91.5\% is achieved. This network layer size was chosen to provide a reasonable trade-off between accuracy and hardware resource requirements.

\subsection{FPGA Implementation Results}

\label{sec:SPAD/res/FPGA}
With the accuracy results provided in the preceding section, we now look at the FPGA implementation results for an instance of the event-based processor whose classification accuracy results were detailed in Figure \ref{fig:NeuronClassificationResults}(c) and (d) for the First-AND event streams and (g) and (h) for the OOBU event streams. Since identical fixed-point implementations were used in both the software and FPGA implementations, the classification accuracy results from the preceding sections apply directly to the end-to-end FPGA implemented system which here is referred to as the NEURO\_NET.

As can be observed from Table \ref{tab:fpgaResource}, a substantial amount of the AER\_REPLAY and NEURO\_NET nodes each consume a significant amount of memory. Ultimately the AER\_REPLAY node would be removed when the SPAD imager is interfacing the event-based processor directly. The memory consumed by the NEURO\_NET node is primarily due to the FIFO that is used for logging purposes. The logging node provides a means of both testing/debugging the neural network design but more importantly allows retrieval of the classification data from the FPGA and into the software. Excluding these hardware costs, the $128\times128$ pixel system requires $125\times125\times4 = 62,500$ addresses with 32 bit DWORD registers which results in a requirement of 250~Kbytes of RAM.

Besides the memory resources consumed, the NEURO\_NET also consumes 128 DSP elements. These DSP elements exist in the NEURO\_CLASSIFICATION node where the vector dot product operation on the output feature map or histogram and the classifier weights occur. Currently, the classifier weights are restricted to a 12-bit signed notation, hence each neuron consumes two embedded 9-bit multipliers (multiplication is pipelined over two clock cycles), for a total of 16 neurons 32 multiplications and with four event polarities the results in a total of 128 multipliers.

\begin{table}
 \centering
 \caption{\textbf{NEURO\_NET hardware resource costs.}}
 \includegraphics[width=.5\textwidth]{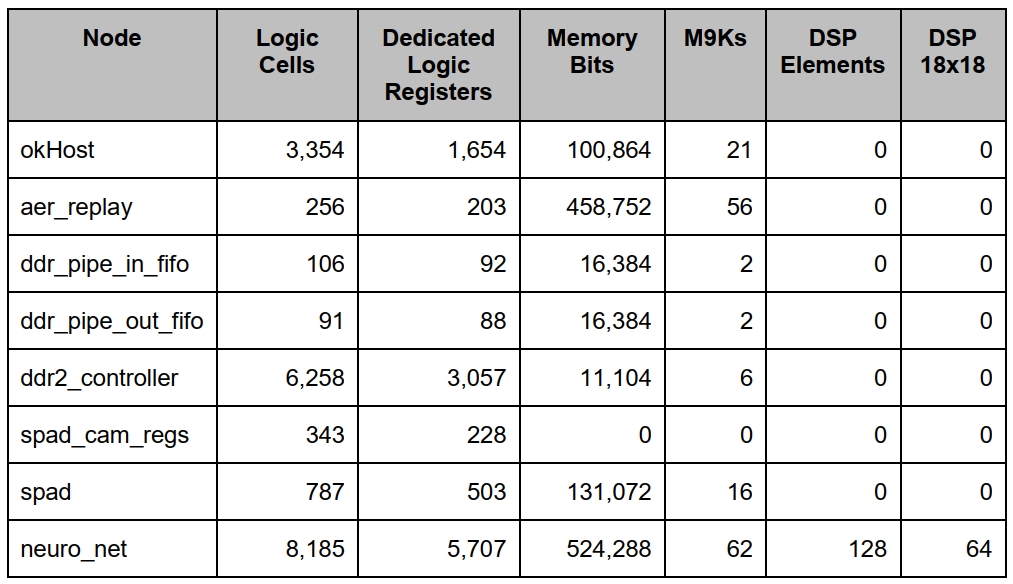}
 
  \label{tab:fpgaResource}
\end{table}

Table \ref{tab:fpgaTiming} lists the execution times for the Neuro Net states, and the timing information has been generated from various Signal tap captures. The total time to process a single AER event and update the feature map or histogram is $~29.24\mu$s. The vast majority of this time is consumed in the 'UPDATE\_PATCH\_MATRICES' state where the ROI from the time surface (or patch data) is read from the DDR2 SDRAM. In this state a total of 100 memory addresses are read, with a patch size of 5, and with four RFs giving $5\times5\times4=100$. On average to synchronize and read a single address of the DDR2 SDRAM from the 100 MHz FPGA clock domain, takes approximately 290 ns, 29 clock cycles. Any future implementations of this event-based processor must take into account this bottle neck. Solutions to this bottleneck include refinement of the logic, faster RAM and finally the development of a cache system outlined in the discussion section.

\begin{table}
 \centering
 \caption{\textbf{NEURO\_NET States Execution Time.}}
 \includegraphics[width=.3\textwidth]{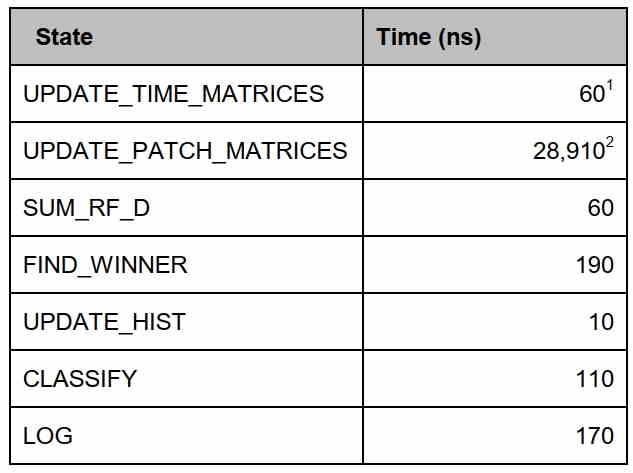}
  \label{tab:fpgaTiming}
\end{table}

\section{Discussion and Future Work}

\label{sec:SPAD/discussion}

A potential weakness in the First-AND event generation method is the all or none behavior of the AND gates whereby even a single faulty, inactive pixel can disable the entire receptive field. While such 'dead pixels' were not observed in the recorded dataset, their potential presence in imagers with higher numbers of pixels is more likely. To protect against the effect of such non-idealities, the replacement of the AND gates with thresholded current summers will be investigated in future designs of the First-AND system.

In order to reduce the potential effect of inherent internal delays in the SPAD array, the global clock may be slowed such that the SPAD pixels are effectively sampled at a lower temporal resolution. In this way, small errors in timing measurement (but also timing measurements of nearby objects) are effectively quantized at the same low-resolution clock cycle. The utility or otherwise of this approach will be investigated in future work.

Given the simple method of their generation and orientation and shape invariant informational properties, OOBU events provide a promising approach to increasing event information in a wide range of event-based sensor systems. The feasibility of implementing these higher level features in local sensor networks will be investigated in future work.
 
As detailed in \ref{sec:SPAD/res/FPGA} a major bottleneck operation of the event-based FPGA processor is the loading of a local ROI from the time surface surrounding a current event. Furthermore, this bottleneck becomes more significant as event polarities and ROI sizes increase. More event polarities are needed when implementing deep event-based convolutional networks and the use of larger ROI sizes can often be beneficial in applications where the underlying signal SNR is low as in event-based space imaging as discussed in \cite{afshar2019star}.
 
While the direct approach to speeding up the ROI read operation is to use faster RAM, other cache-like architectures may provide a solution to the memory loading bottleneck by taking advantage of the likely proximity in space of new events relative to previous ones. In this approach, a slightly larger local region than the ROI may be loaded from RAM to local registers and with each new event, this locally stored address space is first interrogated and if the newest event is close to a previous one, its ROI will have been stored locally and can be fetched at speed. If we assume spatial proximity between temporally proximal events, such an approach is likely to significantly reduce the memory bottleneck associated with the ROI retrieval. In future work, we will investigate potential design solutions to this problem with the aim of providing better timing performance for larger, deeper event-based networks in hardware.
 
\section{Conclusion}
In this work, a challenging SPAD imaging recognition dataset was presented. Three event-based processing approaches were proposed: First-AND, On-Off and On-Off-Bi-polar-Uni-Polar (OOBU) events. The classification accuracy of these event-based methods was investigated and compared to the original frame-based dataset using either linear classifiers or feature extractor networks followed by a linear classifier as a processor. Across all tests, the event-based methods outperform their frame-based counterparts in terms of accuracy and reduced data-rate. This is because the event generation methods involve pooling of raw sensor data over either time or space or both, significantly increasing the information content of each event in comparison to the raw pixel range values in the frame-based data. Within the event-based approaches, the OOBU events resulted in the highest recognition accuracy followed by First-And and On-Off events. In terms of data-rates the First-AND events result in the lowest data-rate followed by the On-Off and OOBU events resulting in 81, 57 and 25 fold reduction in data-rate respectively. In addition to the event-based generation methods, a range of different network parameters were investigated with larger networks shown to outperform smaller ones, trained networks outperforming random networks and two-dimensional spatial pooling outperforming one-dimensional pooling.

The systems investigated in this work provide a range of well performing points in the event-based design space which can be integrated with Direct Time of Flight SPAD sensor hardware to provide event-based processing that not only drastically reduces data-rate coming off the sensor, but also the quality of the output data as it relates to challenging tasks such as a view-invariant classification of large complex datasets. By using the same learning methodology and the same single-layer network structure and by testing across multiple design dimensions such as pooling and network size, we demonstrate exhaustively that the event-based methods outperform the frame-based system across all parameters while serving as a guide for the design of such networks in hardware.
The FPGA implementation of the event-based processor demonstrates the hardware efficiency and processing speed of the design for real-time applications of SPAD sensor technology.
\label{sec:SPAD/conclusion}

\bibliographystyle{IEEEtran}
\bibliography{bare_jrnl.bib}

\begin{IEEEbiography}[{\includegraphics[width=1in,height=1.25in,clip,keepaspectratio]{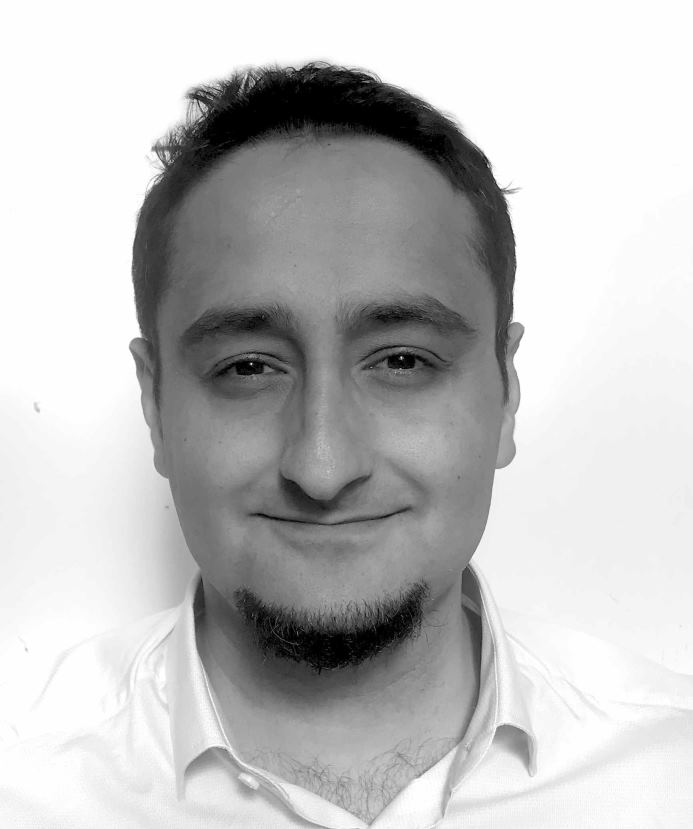}}]{Saeed Afshar} received the B.Eng. degree in electrical engineering from the University of New South Wales, Sydney, Australia, the M.Eng. degree in electrical engineering from
the University of Western Sydney, Sydney, Australia. He is a post-doctoral research fellow with the International Centre for Neuromorphic Systems, at Western Sydney University, Sydney, Australia. His research focuses on event-based vision and audio processing for neuromorphic hardware.
\end{IEEEbiography}

\begin{IEEEbiography}[{\includegraphics[width=1in,height=1.25in,clip,keepaspectratio]{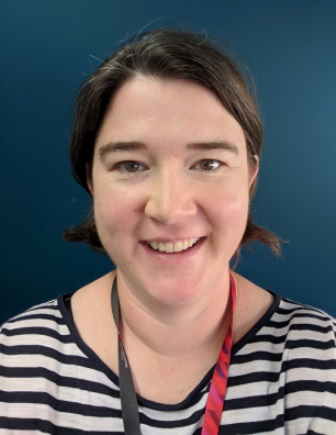}}]{Tara Julia Hamilton} (S’97–M’00) received the B.E. degree (Hons. I) in Electrical Engineering and the B.Com. degree from the University of Sydney, Australia, in 2001, the M.Sc. degree in Biomedical Engineering from the University of New South Wales, Australia, in 2003, and the Ph.D. degree from the University of Sydney in 2009. She is currently an Associate Professor in the School of Engineering at Macquarie University, Australia. Tara has authored over 100 journal papers, conference papers, book chapters, and patents in integrated circuit design, neuromorphic systems, and biomedical engineering. Tara has worked extensively with industry and pursues industrially-relevant research.

\end{IEEEbiography}
\begin{IEEEbiography}[{\includegraphics[width=1in,height=1.25in,clip,keepaspectratio]{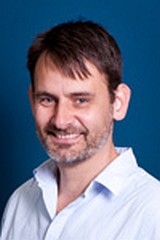}}]{André van Schaik} is a pioneer in Neuromorphic Engineering and the director of the International Centre for Neuromorphic Systems at Western Sydney University. He received his M.Sc. degree in electrical engineering from the University of Twente, Enschede, The Netherlands, in 1990 and his Ph.D. degree in neuromorphic engineering from the Swiss Federal Institute of Technology (EPFL), Lausanne, Switzerland, in 1998. From 1991 until 1994 he was a researcher at the Swiss Centre for Electronics and Microtechnology (CSEM), where he developed the first commercial neuromorphic chip – the optical motion detector used in Logitech trackballs since 1994. In 1998 he was a postdoctoral research fellow in the Department of Physiology at the University of Sydney and in 1999 he became a Senior Lecturer in their School of Electrical and Information Engineering and a Reader in 2004. In 2011 he became a full Professor at Western Sydney University.
His research focuses on all aspects of neuromorphic engineering, encompassing neurophysiology, computational neuroscience, software and algorithm development, and electronic hardware design. He is a Fellow of the IEEE for contributions to Neuromorphic Circuits and Systems. He has authored more than 200 papers and is an inventor of more than 35 patents. He has founded three technology start-ups.

\end{IEEEbiography}

\begin{IEEEbiography}[{\includegraphics[width=1in,height=1.25in,clip,keepaspectratio]{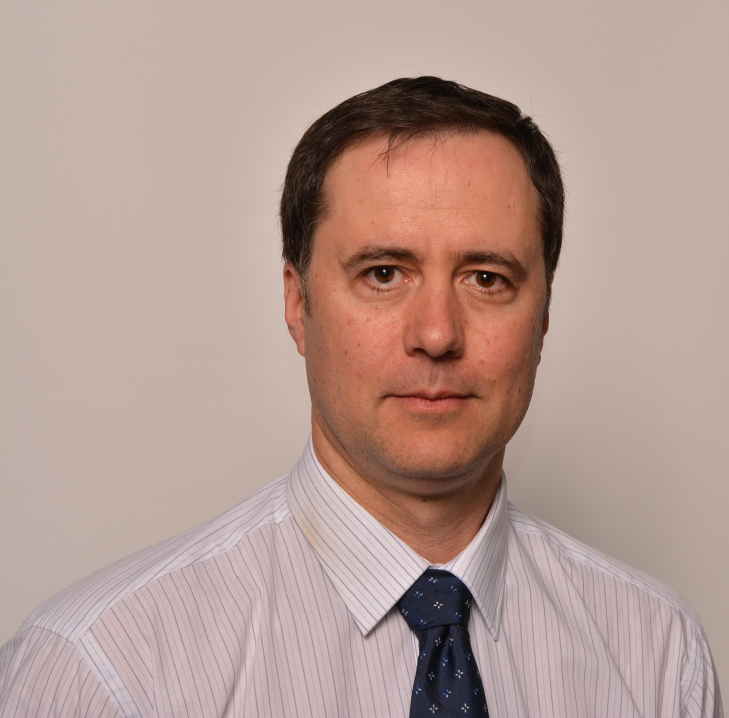}}]{Dennis Delic}
 received the B.E. degree (honours) in electronic engineering and the Ph.D. degree from the University of South Australia, Adelaide, SA, Australia, in 1993 and 2002, respectively.   From 1996 to 2000, he was a Senior Bipolar Design Engineer with Philips Semiconductors, where he was responsible for the design of data interface and power control Application-Specific Integrated Circuits (ASICs). From 2001 to 2006, he was a Senior Designer with Integrated Device Technologies (IDT), designing advanced CMOS Integrated Circuit (IC) programmable switching and timing products. In 2006, he joined the Defence Science and Technology (DST) Group, where he is currently the Emerging Sensors discipline lead and is also an Adjunct Associate Professor at Western Sydney University.  At DST he leads the scientific research of Single-Photon Avalanche Diode (SPAD) detectors and manages a broad portfolio of projects and collaborative agreements with both Industry and University partners to develop the next generation of single photon imaging and recognition technologies for Defence applications.   His current research interests include  IC Design and semiconductor sensing technologies, advanced single photon imagers which include Single Photon LiDAR (SPL) for both bathymetric and 3-D imaging applications.  Dr. Delic received the DST Fellowship to research SPAD image sensors in 2012.
\end{IEEEbiography}

\end{document}